\documentclass[oneside,english]{amsart}
\usepackage{lmodern}

\usepackage[T1]{fontenc}
\usepackage[latin9]{inputenc}
\usepackage{color}
\usepackage{babel}
\usepackage{amsthm}
\usepackage{amssymb}
\usepackage[all]{xy}
\usepackage[unicode=true,
 bookmarks=true,bookmarksnumbered=false,bookmarksopen=false,
 breaklinks=false,pdfborder={0 0 1},backref=false,colorlinks=true]
 {hyperref}
\hypersetup{pdftitle={A Gamma Class Formula for Open Gromov-Witten Calculations},
 pdfauthor={Matthew Mahowald},
 pdfsubject={Gromov-Witten theory},
 pdfkeywords={Gromov-Witten theory, enumerative geometry, stable maps, A-model}}

\makeatletter
  \theoremstyle{plain}
  \newtheorem*{thm*}{\protect\theoremname}
\theoremstyle{plain}
\newtheorem{thm}{\protect\theoremname}
  \theoremstyle{remark}
  \newtheorem*{rem*}{\protect\remarkname}
  \theoremstyle{plain}
  \newtheorem{cor}[thm]{\protect\corollaryname}
  \theoremstyle{plain}
  \newtheorem{lem}[thm]{\protect\lemmaname}

\usepackage{tikz}
\usetikzlibrary{matrix,arrows}

\usepackage{import}

\newcommand{\Def}{\mathrm{Def}}
\newcommand{\Aut}{\mathrm{Aut}}
\newcommand{\Obs}{\mathrm{Obs}}
\newcommand{\Ext}{\mathrm{Ext}}

\newcommand{\vdim}{\mathrm{vdim}~}

\newcommand{\rank}{\mathrm{rank}~}

\newcommand{\arxiv}[2]{\href{http://arxiv.org/abs/#1}{\texttt{arXiv:#1 #2}}}

\makeatother

  \providecommand{\corollaryname}{Corollary}
  \providecommand{\lemmaname}{Lemma}
  \providecommand{\remarkname}{Remark}
  \providecommand{\theoremname}{Theorem}
\providecommand{\theoremname}{Theorem}

\begin{document}

\title{A Gamma Class Formula for Open Gromov-Witten Calculations}

\author{}

\author{Matthew Mahowald}

\address{Matthew~Mahowald, Department~of~Mathematics, Northwestern~University,
Evanston,~IL 60208, USA}

\email{mamahowa@math.northwestern.edu}
\begin{abstract}
For toric Calabi-Yau 3-folds, open Gromov-Witten invariants associated
to Riemann surfaces with one boundary component can be written as
the product of a disk factor and a closed invariant. Using the Brini-Cavalieri-Ross
formalism, these disk factors can often be expressed in terms of gamma
classes. When the Lagrangian boundary cycle is preserved by the torus
action and can be locally described as the fixed locus of an anti-holomorphic
involution, we prove a formula that expresses the disk factor in terms
of a gamma class and combinatorial data about the image of the Lagrangian
cycle in the moment polytope. We verify that this formula encodes
the expected invariants obtained from localization by comparing with
several examples. We then examine a novel application of this formula
to disk enumeration on the quintic 3-fold. Finally, motivated by large
$N$ duality, we show that this formula also unexpectedly applies
to Lagrangian cycles on $\mathcal{O}_{\mathbb{P}^{1}}(-1,-1)$ constructed
from torus knots. 
\end{abstract}
\maketitle
\tableofcontents{}

\pagebreak{}

\section{Introduction}

\subsection{Background and motivation}

Gromov-Witten theory has a rich history, both in physics and mathematics.
Physically, Gromov-Witten invariants appear in type IIA topological
string theory as instanton counts associated to interactions between
particles. Mathematically, they are invariants associated to symplectic
manifolds that, roughly speaking, count pseudoholomorphic curves in
the manifold. The relationship between these two perspectives is conceptually
straightforward: as a string moves in time, it sweeps out a compact
Riemann surface (its `worldsheet'). The amplitudes in string theory
encode counts of maps from Riemann surfaces into a 3-(complex)-dimensional
Calabi-Yau manifold, and Gromov-Witten theory assigns invariants to
spaces of such maps. 

In general, counting holomorphic maps from Riemann surfaces to a given
target space is a difficult problem in enumerative geometry. Gromov-Witten
theory has famously benefited from its connections with string dualities,
first with mirror symmetry \cite{COGP,WittenABModels}, and more recently,
large $N$ duality \cite{GopakumarVafa,OoguriVafa}. Beginning with
\cite{Kontsevich}, for toric manifolds, Gromov-Witten invariants
associated to maps of closed surfaces have also been systematically
computed using localization \cite{GraberPandharipande,LMS,KlemmZaslow}.
``Closed'' Gromov-Witten theory is a natural mathematical counterpart
to closed topological string theory, and, in contrast to the ``open''
theory (i.e., for maps of Riemann surfaces with boundary), the moduli
spaces involved are rigorously defined. 

Open Gromov-Witten theory is the subject of this paper. By analogy
with the closed case, open Gromov-Witten theory is a mathematical
counterpart to open topological string theory: open strings sweep
out compact Riemannian surfaces with boundary, and the boundary of
the strings are constrained to lie on branes. These boundary constraints
are expressed mathematically as Lagrangian submanifolds $L\subset X$,
and the string amplitudes are encoded by counts of holomorphic maps
$f:\Sigma\rightarrow X$, with the image of the boundary constrained
to lie on $L$: $f\left(\partial\Sigma\right)\subset L$. However,
as observed in \cite{AKV,KatzLiu}, there are additional subtleties
in adapting the methods of the closed theory to the open case. In
particular, even for well-behaved Lagrangian boundary cycles, open
Gromov-Witten invariants depend on an additional integral parameter
(in localization, this parameter corresponds to the weights of the
torus action). 

In spite of this, the same computational tools of mirror symmetry,
large $N$ duality, and localization can still be used. In fact, through
these string dualities, open Gromov-Witten theory can be connected
to both classical and homological knot theory \cite{OoguriVafa,DSV,GuJockersKlemmSoroush,MarinoVafa,WittenCSThy,WittenCSandHOMFLY}.
Motivated by relationships with the crepant resolution conjecture
\cite{CRC}, open Gromov-Witten theory can also be generalized to
orbifolds \cite{BrCav}. This setting has driven a more abstract formulation
of open Gromov-Witten invariants, which has led to a deeper understanding
of the A-model. In particular, as will be discussed in detail below,
the open invariants contain gamma classes coming from disk terms \cite{BCR-Gamma-fun}.
In addition, the open Gromov-Witten generating function can be obtained
from a modification of Givental's $J$ function \cite{BrCav,Brini-J-fun}. 

The primary goal of this paper is to describe a concise and consistent
framework for computing open Gromov-Witten invariants directly, via
localization. Leveraging the formalism of \cite{BCR-Gamma-fun}, the
main tool is a formula for open Gromov-Witten invariants expressed
in terms of local combinatorial data and a gamma class. As expected
from \cite{AKMV}, the construction depends only on the local geometry
near a vertex of the moment polytope of $X$. In the case where the
associated moduli space of open maps is rigorously defined (\cite{KatzLiu}),
this formula is proven to be correct. Most intriguingly, this result
is shown to apply in two unexpected contexts: enumerative invariants
associated to the quintic 3-fold, and Lagrangian cycles obtained from
torus knots appearing in large $N$ duality.

The author hopes that the approach described herein will lead to a
more general construction of open Gromov-Witten invariants.

\subsection{Organization of the paper}

The paper is organized in the following way. Section~\ref{sec:Preliminaries}
reviews some general facts about open Gromov-Witten theory, including
deformation theory and localization. Most importantly, this section
describes how to express an open Gromov-Witten invariant as the product
of a ``disk term'' and an invariant of closed maps. Section~\ref{sec:Main-Results}
contains the proof of the main computational tool of this paper: 
\begin{thm*}
Let $X$ be a Calabi-Yau 3-fold and $L\subset X$ a Lagrangian submanifold.
Let $S^{1}$ act on $X$ such that the $S^{1}$ action preserves $L$,
and $L$ intersects a rigid circle-invariant curve $C$. Suppose that
$L$ can be described in a neighborhood of $L\cap C$ as the fixed
locus of an anti-holomorphic involution. Let $\gamma\in H^{2}\left(X;\mathbb{Q}\right)$.
Then, the genus $g$, 1 boundary component, degree $d$, winding $w$
open Gromov-Witten invariant with Lagrangian boundary $L$ is
\[
\left\langle \gamma\right\rangle _{d,w}^{g,1}=\left.\left(\Delta_{X,L}\circ\left\langle \gamma,\frac{\phi_{p}}{z-\psi}\right\rangle _{g,d}\right)\right|_{z=\alpha},
\]
where $\Delta_{X,L}$ is the disk function 
\[
\Delta_{X,L}\left(\gamma\right):=\frac{\pi}{wz\widehat{\Gamma}_{X}\sin\left(\pi\frac{\lambda}{z}\right)}\cdot\gamma.
\]
Here, $\widehat{\Gamma}_{X}$ is the homogeneous Iritani gamma class,
$\lambda$ is the weight of the $S^{1}$ action along a normal direction
to $C$, $\alpha=c_{1}(T_{0}\Delta)$ is the equivariant Chern class
of the induced representation of $S^{1}$ at the attachment point
of the disk, and $\phi_{p}$ is the equivariant class of the image
$p\in X$ of the disk attachment point. 
\end{thm*}
This gamma class formula was encountered previously in \cite{BCR-Gamma-fun},
where the authors study a Lagrangian locally described as the fixed
locus of the antiholomorphic involution $\sigma\left(\xi,x,y\right)=\left(1/\overline{\xi},\overline{y\xi},\overline{x\xi}\right)$.
As will be seen (lemma \ref{lem:involution-form}), this result implies
the formula above after a change of coordinates. Section~\ref{sec:Comparisons}
describes how to apply the gamma class formula to several examples
where the resulting invariant is already known, and demonstrates that
this formula reproduces the expected result. The last two sections
study two examples where the assumptions on the local geometry of
$L$ are not satisfied. Section~\ref{sec:quintic} examines disk
enumeration on the quintic 3-fold and finds that a slight modification
of this formula again applies. Finally, Section~\ref{sec:torus-knot-lagrangians}
applies this formula to a novel class of Lagrangian cycles motivated
by large $N$ duality. These Lagrangian cycles are obtained from the
conormal bundles of torus knots in $S^{3}$ after the conifold transition,
and do not have the same local description required in the above theorem.
Nevertheless, the main result of this paper is still found to apply
to these cycles. The examples in sections \ref{sec:quintic} and \ref{sec:torus-knot-lagrangians}
hint that a version of the theorem may hold for a broader class of
Lagrangian cycles.

\subsection*{Acknowledgments}

The author thanks H.~Gao, H.~Jockers, C-C.~M.~Liu, and P.~Zhou
for valuable discussions, and A.~Brini for helpful corrections. The
author is especially grateful to R.~Cavalieri for the suggestion
of this project and many related conversations, and to E.~Zaslow
for guidance and suggestions.

\section{Preliminaries\label{sec:Preliminaries}}

\subsection{Deformation theory for stable maps\label{sub:stable-open-maps}}

Open Gromov-Witten invariants are enumerative invariants of maps $f:\Sigma\rightarrow X$
from Riemann surfaces with boundary into a Calabi-Yau manifold $X$
with a chosen Lagrangian submanifold $L$, such that $f\left(\partial\Sigma\right)\subset L$.
By analogy with the definition for closed stable maps appearing in
\cite{Kontsevich}, \cite{KatzLiu} define the open Gromov-Witten
invariant in the following way. Fix integers $g$ (the genus of $\Sigma$)
and $h$ (the number of connected components of $\partial\Sigma$),
and a relative homology class $d\in H_{2}\left(X,L;\mathbb{Z}\right)$
with $\partial d=\sum w_{i}\in H_{1}\left(L;\mathbb{Z}\right)$. Then,
the open Gromov-Witten invariant $GW_{d,w_{1},\ldots,w_{h}}^{g,h}$
is a virtual count of continuous maps $f:\left(\Sigma,\partial\Sigma\right)\rightarrow\left(X,L\right)$
satisfying:
\begin{itemize}
\item $\left(\Sigma,\partial\Sigma\right)$ is a Riemann surface of genus
$g$ with boundary $ $$\partial\Sigma$ consisting of $h$ oriented
circles,
\item $f$ is holomorphic in the interior of $\Sigma$,
\item $f_{*}\left[\Sigma\right]=d$, and
\item $f_{*}\left[\partial\Sigma\right]=\sum w_{i}$. 
\end{itemize}
For brevity, $w_{1},\ldots,w_{h}$ will sometimes be denoted by $\overrightarrow{w}$.
In order to define such an invariant, \cite{KatzLiu} construct a
moduli space $\overline{\mathcal{M}}_{g,h}\left(X,L;d,\overrightarrow{w}\right)$
of stable maps which compactify the maps described above, and give
a local description of an orientation and a virtual fundamental class
on this moduli space. In particular, the authors generalize the deformation
complex in ordinary Gromov-Witten theory to the open case.

Recall that for smooth, closed $\Sigma$ in ordinary Gromov-Witten
theory, there is a normal bundle exact sequence of vector bundles
on $\Sigma$: 
\[
\xymatrix{0\ar[r] & T_{\Sigma}\ar[r] & f^{*}T_{X}\ar[r] & N_{\Sigma/X}\ar[r] & 0}
.
\]
The corresponding long exact sequence in cohomology is \begin{center}
\begin{tikzpicture}[descr/.style={fill=white,inner sep=1.5pt}]
        \matrix (m) [
            matrix of math nodes,
            row sep=1em,
            column sep=2.5em,
            text height=1.5ex, text depth=0.25ex
        ]
        { 0 & H^0(\Sigma,T_{\Sigma}) & H^0(\Sigma, f^{*}T_{X}) & H^0(\Sigma, N_{\Sigma/X}) & \\
            & H^1(\Sigma,T_{\Sigma}) & H^1(\Sigma, f^{*}T_{X}) & H^1(\Sigma, N_{\Sigma/X}) & 0 \\
        };

        \path[overlay,->, font=\scriptsize,>=latex]
        (m-1-1) edge (m-1-2)
        (m-1-2) edge (m-1-3)
        (m-1-3) edge (m-1-4)
        (m-1-4) edge[out=355,in=175] (m-2-2)
        (m-2-2) edge (m-2-3)
        (m-2-3) edge (m-2-4)
        (m-2-4) edge (m-2-5);
\end{tikzpicture}
\end{center} The terms in this sequence can be interpreted as infinitesimal automorphisms,
deformations, and obstructions to deformations for $\Sigma$ and $f$,
so this sequence can be re-written as the deformation complex: \begin{equation}
\begin{tikzpicture}[descr/.style={fill=white,inner sep=1.5pt}, baseline=(current bounding box).center)]
        \matrix (m) [
            matrix of math nodes,
            row sep=1em,
            column sep=2.5em,
            text height=1.5ex, text depth=0.25ex
        ]
        { 0 & \Aut(\Sigma) & \Def(f) & \Def(\Sigma,f) & \\
            & \Def(\Sigma) & \Obs(f) & \Obs(\Sigma,f) & 0. \\
        };

        \path[overlay,->, font=\scriptsize,>=latex]
        (m-1-1) edge (m-1-2)
        (m-1-2) edge (m-1-3)
        (m-1-3) edge (m-1-4)
        (m-1-4) edge[out=355,in=175] (m-2-2)
        (m-2-2) edge (m-2-3)
        (m-2-3) edge (m-2-4)
        (m-2-4) edge (m-2-5);
\end{tikzpicture}
\label{eq:deformation-complex}
\end{equation} Suitably interpreted, the same sequence holds for nodal, open curves
in open Gromov-Witten theory: Over a smooth point $\left(\Sigma,f\right)$
in $\overline{\mathcal{M}}_{g,h}\left(X,L;d,\overrightarrow{w}\right)$,
$H^{k}\left(\Sigma,f^{*}T_{X}\right)$ are the cohomology groups associated
to sections $s$ of $\left(\Sigma,f^{*}T_{X}\right)$ satisfying $s|_{\partial\Sigma}\in\Gamma\left(\partial\Sigma,f^{*}T_{L}\right)$. 

The expected (virtual) dimension of $\overline{\mathcal{M}}_{g,h}\left(X,L;d,\overrightarrow{w}\right)$
is 
\begin{eqnarray*}
\vdim\overline{\mathcal{M}}_{g,h}\left(X,L;d,\overrightarrow{w}\right) & = & \rank\Def\left(\Sigma,f\right)-\rank\Obs\left(\Sigma,f\right)\\
 & = & \mu\left(f^{*}T_{X},f|_{\partial\Sigma}^{*}T_{L}\right)-\left(\dim X-3\right)\chi\left(\Sigma\right),
\end{eqnarray*}
where $\mu$ denotes the generalized Maslov index of the real subbundle
$\left(f|_{\partial\Sigma}\right)^{*}T_{L}\subset f^{*}T_{X}$ \cite{KatzLiu}.
When $X$ is a complex manifold and $L$ is the fixed locus of an
anti-holomorphic involution, $\mu\left(f^{*}T_{X},\left(f|_{\partial\Sigma}\right)^{*}T_{L}\right)=\int_{d}c_{1}\left(T_{X}\right)$,
and if $X$ is a Calabi-Yau threefold, $\vdim\overline{\mathcal{M}}_{g,h}\left(X,L;d,\overrightarrow{w}\right)=0$.
When $\overline{\mathcal{M}}_{g,h}\left(X,L;d,\overrightarrow{w}\right)$
has a well-behaved torus action, \cite{KatzLiu} give an explicit
description for the localization of the virtual fundamental class
to the fixed loci of the torus action. In contrast to closed Gromov-Witten
theory, the virtual cycle found in \cite{KatzLiu} depends on the
torus action. Additionally, the invariants defined in \cite{KatzLiu}
depend on a choice of orientation. This choice is reflected in the
overall sign of the invariant, and the invariant formula proposed
in this note has an analogous orientation-dependent sign.

\subsection{Separating the disk term\label{sub:Separating-the-disc}}

Gromov-Witten invariants are, in general, difficult to compute. The
primary computational tool is the Atiyah-Bott fixed-point formula
\cite{AtiyahBott}. When applied to computations of Gromov-Witten
invariants for toric varieties, this ``localizes'' integrals over
the entire moduli space of stable maps to integrals over only those
maps which are fixed by the torus action \cite{Kontsevich,GraberPandharipande,KlemmZaslow}.
The Atiyah-Bott fixed point formula for the integral of a class $\phi$
over a manifold (or more generally, a Deligne-Mumford stack) $M$
is 
\begin{equation}
\int_{M}\phi=\sum_{P}\int_{P}\left(\frac{i_{P}^{*}\phi}{e\left(N_{P}\right)}\right),\label{eq:atiyah-bott-localization}
\end{equation}
where the sum is over the fixed point sets $P$, $i_{P}$ is the embedding
of $P$ into $M$, and $e\left(N_{P}\right)$ is the (equivariant)
Euler class of the normal bundle of $P$ in $M$. 

Following \cite{GraberPandharipande,Kontsevich}, a stable map $\left(\Sigma,f\right)$
can be naturally described as a decorated graph. The vertices $v$
of the graph correspond to contracted components of the nodal curve
$\Sigma$, and are labeled by the genus $g\left(v\right)$ of that
component. The edges correspond to $\mathbb{P}^{1}$'s which are not
contracted by $f$, and are labeled by the degree $d_{e}$ of the
map $f|_{\mathbb{P}^{1}}$ associated to the edge. When the fixed
stable maps are described as decorated graphs in this way, \eqref{eq:atiyah-bott-localization}
becomes 
\begin{equation}
GW_{d}^{g}:=\int_{\left[\overline{\mathcal{M}}_{g,0}\left(X,d\right)\right]^{vir}}1=\sum_{\Gamma}\frac{1}{\left|A_{\Gamma}\right|}\int_{M_{\Gamma}}\frac{1}{e\left(N_{\Gamma}^{vir}\right)}.\label{eq:localization-for-closed-gw-invt}
\end{equation}

As observed in \cite{GraberZaslow}, the graph description of stable
maps can be extended to the open stable maps defined in \cite{KatzLiu}
by treating the open disk component as a ``leg'' of the graph. A
crucial consequence of this is that open Gromov-Witten invariants
can be expressed as a closed Gromov-Witten invariant multiplied by
a ``disk term.'' For simplicity, restrict attention to surfaces
with one boundary component. Let $X$ be a Calabi-Yau manifold equipped
with an $S^{1}$ action that fixes a Lagrangian submanifold $L\subset X$.
Suppose that $f\colon\Sigma\rightarrow X$ is a stable map from a
genus $g$ Riemann surface with one boundary component such that $f_{*}\left[\Sigma\right]=d\in H_{2}\left(X;\mathbb{Z}\right)$
and $f|_{\partial\Sigma}\colon\partial\Sigma\rightarrow f\left(\partial\Sigma\right)$
has winding $w$ as a map between homotopy circles. Let $\overline{\mathcal{M}}:=\overline{\mathcal{M}}_{g,1,0}\left(X,L;d,w\right)$
denote the the moduli space of such maps (genus $g$, 1 boundary component,
$0$ marked points).

The $S^{1}$ action on $X$ naturally induces an $S^{1}$ action on
$\overline{\mathcal{M}}$. If $\left(\Sigma,f\right)\in\overline{\mathcal{M}}$
is fixed by the $S^{1}$ action, then $\Sigma$ must take the form
\[
\Sigma=\Sigma_{0}\cup_{\nu}\Delta,
\]
where $\Sigma_{0}$ is a closed genus $g$ Riemann surface, $\Delta$
is a disk, and $\nu$ is a simple node on $\Sigma_{0}$ at which $\Delta$
is attached. $S^{1}$ invariance further requires that $\left(\Sigma_{0},f|_{\Sigma_{0}},\nu\right)$
is fixed by the induced action on $\overline{\mathcal{M}}_{g,1}\left(X,d\right)$. 

Then, a virtual localization formula analogous to \eqref{eq:localization-for-closed-gw-invt}
for the genus $g$, degree $d$, winding $w$ open Gromov-Witten invariant
would take the form 
\begin{equation}
GW_{d,w}^{g,1}=\int_{\overline{\mathcal{M}}^{vir}}1=\sum_{\Gamma}\frac{1}{\left|\mathbf{A}_{\Gamma}\right|}\int_{M_{\Gamma}}\frac{1}{e\left(N_{\Gamma}^{vir}\right)},\label{eq:localization-for-open-gw-invt}
\end{equation}
where $\mathbf{A}_{\Gamma}=\mathbb{Z}/w\mathbb{Z}\times A_{\Gamma'}$
($\Gamma'$ is the graph associated to the closed curve $\left(\Sigma_{0},\nu\right)$). 

Note that $\overline{\mathcal{M}}_{g,1}\left(X,d\right)$ is equipped
with a natural map $e{}_{\nu}:\overline{\mathcal{M}}_{g,1}\left(X,d\right)\rightarrow X$
given by evaluation at the marked point. The conditions on $\left(\Sigma,f\right)$
specified above (in particular, that $f\left(\nu\right)=p$) imply
that the fixed locus $M_{\Gamma}$ is isomorphic to the fixed subspace
$e{}_{\nu}^{-1}(p)^{S^{1}}\subset\overline{\mathcal{M}}_{g,1}\left(X,d\right)^{S^{1}}$. 

As in the closed case, the equivariant normal bundle $e\left(N_{\Gamma}^{vir}\right)$
and the virtual fundamental cycle are determined, respectively, by
the moving and fixed parts of the deformation complex \eqref{eq:deformation-complex}:
\begin{center}
\begin{tikzpicture}[descr/.style={fill=white,inner sep=1.5pt}]
        \matrix (m) [
            matrix of math nodes,
            row sep=1em,
            column sep=2.5em,
            text height=1.5ex, text depth=0.25ex
        ]
        { 0 & \Aut(\Sigma) & \Def(f) & \Def(\Sigma,f) & \\
            & \Def(\Sigma) & \Obs(f) & \Obs(\Sigma,f) & 0 \\
        };

        \path[overlay,->, font=\scriptsize,>=latex]
        (m-1-1) edge (m-1-2)
        (m-1-2) edge (m-1-3)
        (m-1-3) edge (m-1-4)
        (m-1-4) edge[out=355,in=175] (m-2-2)
        (m-2-2) edge (m-2-3)
        (m-2-3) edge (m-2-4)
        (m-2-4) edge (m-2-5);
\end{tikzpicture}
\end{center} This gives the following relationship in the representation ring
of $S^{1}$: 
\[
\Obs\left(\Sigma,f\right)-\Def\left(\Sigma,f\right)=\Aut\left(\Sigma\right)+\Obs\left(f\right)-\Def\left(\Sigma\right)-\Def\left(f\right)
\]
with $\Obs\left(f\right)=H^{1}\left(\Sigma,f^{*}T_{X}\right)$, $\Def\left(f\right)=H^{0}\left(\Sigma,f^{*}T_{X}\right)$,
$\Aut\left(\Sigma\right)=\Ext^{0}\left(\Omega_{\Sigma}\left(D\right),\mathcal{O}_{\Sigma}\right)$,
and $\Def\left(\Sigma\right)=\Ext^{1}\left(\Omega_{\Sigma}\left(D\right),\mathcal{O}_{\Sigma}\right)$.
(Here, $D$ is the divisor associated to the nodal points of $\Sigma$.
When $\Sigma$ is smooth, these spaces are just $H^{0}\left(\Sigma,T_{\Sigma}\right)$
and $H^{1}\left(\Sigma,T_{\Sigma}\right)$, respectively). 

Now, relate the terms in this sequence to the terms concerning $\Sigma_{0}$
and $\Delta$: Let $f_{0}:=f|_{\Sigma_{0}}$ and $f_{\Delta}:=f|_{\Delta}$.
Suppose that $\Delta$ is parametrized by $\left\{ \left|t\right|\leq1\right\} $,
with $\nu$ identified with the point $t=0$. Then, there is an exact
sequence \begin{center}
\begin{tikzpicture}[descr/.style={fill=white,inner sep=1.5pt}]
        \matrix (m) [
            matrix of math nodes,
            row sep=1em,
            column sep=2.5em,
            text height=1.5ex, text depth=0.25ex
        ]
        { 0 & \mathcal{O}_{\Sigma} & \mathcal{O}_{\Sigma_{0}} \oplus \mathcal{O}_{\Delta} & \mathcal{O}_{\nu} & 0.\\
        };

        \path[overlay,->, font=\scriptsize,>=latex]
        (m-1-1) edge (m-1-2)
        (m-1-2) edge (m-1-3)
        (m-1-3) edge (m-1-4)
        (m-1-4) edge (m-1-5);
\end{tikzpicture}
\end{center} This becomes the exact sequence on cohomology: \begin{center}
\begin{tikzpicture}[descr/.style={fill=white,inner sep=1.5pt}]
        \matrix (m) [
            matrix of math nodes,
            row sep=1em,
            column sep=2.5em,
            text height=1.5ex, text depth=0.25ex
        ]
        { 0 & \Def(f) & H^{0}(\Delta,T_{(\Delta,f_{\Delta})}) \oplus \Def(f_{0}) & T_{p}X & \\
            & \Obs(f) & H^{1}(\Delta,T_{(\Delta,f_{\Delta})}) \oplus \Obs(f_{0}) & 0 &  \\
        };

        \path[overlay,->, font=\scriptsize,>=latex]
        (m-1-1) edge (m-1-2)
        (m-1-2) edge (m-1-3)
        (m-1-3) edge (m-1-4)
        (m-1-4) edge[out=355,in=175] (m-2-2)
        (m-2-2) edge (m-2-3)
        (m-2-3) edge (m-2-4);
\end{tikzpicture}
\end{center} which yields the following relations in the representation ring:
\begin{align*}
\Obs(f)^{f}-\Def(f)^{f} & =H^{1}(\Delta,T_{(\Delta,f_{\Delta})})^{f}-H^{0}(\Delta,T_{(\Delta,f_{\Delta})})^{f}\\
 & +\Obs(f_{0})^{f}-\Def(f_{0})^{f},
\end{align*}
\begin{align*}
\Obs(f)^{m}-\Def(f)^{m} & =H^{1}(\Delta,T_{(\Delta,f_{\Delta})})^{m}-H^{0}(\Delta,T_{(\Delta,f_{\Delta})})^{m}\\
 & +\Obs(f_{0})^{m}-\Def(f_{0})^{m}+T_{p}X,
\end{align*}
where $p=f\left(\nu\right)\in X$ and the $f$, $m$ superscripts
denote fixed and moving terms with respect to the $S^{1}$ action. 

Similarly, 
\begin{eqnarray*}
\Aut(\Sigma)^{m} & = & \Aut\left(\Sigma_{0},\nu\right)^{m}+\Aut\left(\Delta,0\right)^{m},\\
\Aut(\Sigma)^{f} & = & \Aut\left(\Sigma_{0},\nu\right)^{f}+\Aut\left(\Delta,0\right)^{f},\\
\Def\left(\Sigma\right)^{f} & = & \Def\left(\Sigma_{0},\nu\right)^{f},
\end{eqnarray*}
and 
\[
\Def\left(\Sigma\right)^{m}=\Def\left(\Sigma_{0},\nu\right)^{m}+T_{\nu}\Sigma_{0}\otimes T_{0}\Delta.
\]
Note that $\Aut\left(\Delta,0\right)$ consists of the infinitesimal
automorphisms of $\Delta$ preserving the origin $t=0$, which are
generated by the sections $t\partial_{t}$ over $\mathbb{R}$. Therefore,
$\Aut\left(\Delta,0\right)^{m}$ is trivial, and $\Aut\left(\Delta,0\right)^{f}=\mathbb{R}$. 

Collecting the above observations, 
\begin{align*}
\Obs(\Sigma,f)^{f}-\Def\left(\Sigma,f\right)^{f} & =H^{1}(\Delta,T_{(\Delta,f_{\Delta})})^{f}-H^{0}(\Delta,T_{(\Delta,f_{\Delta})})^{f}\\
 & +\Obs(f_{0})^{f}-\Def(f_{0})^{f}\\
 & +\Aut(\Sigma_{0},\nu)^{f}-\Def(\Sigma_{0},\nu)^{f}\\
 & +\Aut(\Delta,0)^{f}
\end{align*}
and 
\begin{align*}
\Obs(\Sigma,f)^{m}-\Def\left(\Sigma,f\right)^{m} & =H^{1}(\Delta,T_{(\Delta,f_{\Delta})})^{m}-H^{0}(\Delta,T_{(\Delta,f_{\Delta})})^{m}\\
 & +\Obs(f_{0})^{m}-\Def(f_{0})^{m}\\
 & +\Aut(\Sigma_{0},\nu)^{m}-\Def(\Sigma_{0},\nu)^{m}\\
 & +T_{p}X-T_{\nu}\Sigma_{0}\otimes T_{0}\Delta
\end{align*}
The first equation implies that the virtual fundamental cycle of the
fixed locus is the restriction of the natural virtual cycle of the
fixed locus $\left[\overline{\mathcal{M}_{g,1}}\left(X,d\right)^{S^{1}}\right]^{vir}$
to the subspace $e{}_{\nu}^{-1}(p)^{S^{1}}$. The second equation
yields the following relationship between the normal bundles $e(N_{\Gamma}^{vir})$
and $e\left(N_{\Gamma'}^{vir}\right)$ (where $\Gamma'$ is the graph
for the closed curve with the disk ``leg'' removed, i.e., the stable
map associated to $\Sigma_{0}$ with one marked point): 
\begin{align*}
N_{\Gamma}^{vir} & =N_{\Gamma'}^{vir}-T_{p}X+R\mathbb{L}^{-1}\\
 & -H^{1}(\Delta,T_{(\Delta,f_{\Delta})})^{m}+H^{0}\left(\Delta,T_{(\Delta,f_{\Delta})}\right)^{m}.
\end{align*}
Here $R$ is the representation of $S^{1}$ on $T_{0}\Delta\cong\mathbb{C}$
induced by the pullback of the $S^{1}$ action on $f\left(\Delta\right)$,
and $\mathbb{L}$ is the tautological cotangent line bundle on $\overline{\mathcal{M}_{g,1}}\left(X,d\right)$
associated to the marked point $\nu$, i.e., the line bundle whose
fiber at the point $\left(f_{0},\Sigma_{0},\nu\right)$ is $T_{\nu}^{*}\Sigma_{0}$.
$R\mathbb{L}^{-1}$ is contribution from the term $T_{\nu}\Sigma_{0}\otimes T_{0}\Delta$:
$T_{\nu}\Sigma_{0}$ is the fiber of $\mathbb{L}$ and $T_{0}\Delta$
is a constant vector space which carriers the representation $R$
by $S^{1}$. 

Hence, 
\begin{equation}
\int_{M_{\Gamma}}\frac{1}{e\left(N_{\Gamma}^{vir}\right)}=\frac{e_{S^{1}}\left(H^{1}\left(\Delta,T_{(\Delta,f_{\Delta})}\right)\right)e_{S^{1}}\left(T_{p}X\right)}{e_{S^{1}}\left(H^{0}\left(\Delta,T_{(\Delta,f_{\Delta})}\right)\right)}\int_{M_{\Gamma'}}\frac{1}{e\left(N_{\Gamma'}^{vir}\right)\left(\alpha-\psi\right)},\label{eq:sep-of-disc-contribution}
\end{equation}
where $\alpha=c_{1}\left(R\right)$ and $\psi=c_{1}\left(\mathbb{L}\right)$
(so that $e_{S^{1}}\left(R\mathbb{L}^{-1}\right)=\alpha-\psi$). As
before, $e_{S^{1}}\left(\cdot\right)$ denotes the $S^{1}$-equivariant
Euler class of the specified bundle. Denote by $D_{X,L}$ the ``disk
factor'' 
\begin{equation}
D_{X,L}:=\left(\frac{1}{w}\right)\frac{e_{S^{1}}\left(H^{1}\left(\Delta,T_{(\Delta,f_{\Delta})}\right)\right)}{e_{S^{1}}\left(H^{0}\left(\Delta,T_{(\Delta,f_{\Delta})}\right)\right)}.\label{eq:DXL}
\end{equation}
 Then, \eqref{eq:localization-for-open-gw-invt} becomes 
\[
GW_{d,w}^{g}=\int_{\overline{\mathcal{M}}^{vir}}1=D_{X,L}\left(\sum_{\Gamma'}\frac{1}{\left|A_{\Gamma'}\right|}\int_{M_{\Gamma'}}\frac{i^{*}ev^{*}\left(\phi_{p}\right)}{e\left(N_{\Gamma'}^{vir}\right)\left(\alpha-\psi\right)}\right),
\]
where $\phi_{p}$ is the equivariant Thom class of the point $p\in X$,
and $i^{*}$ the pullback to the fixed locus $M_{\Gamma'}$. Comparing
this formula with \eqref{eq:localization-for-closed-gw-invt} shows
that the parenthetical quantity is the localization of a closed Gromov-Witten
invariant: 
\begin{equation}
GW_{d,w}^{g,1}=D_{X,L}\int_{\left[\overline{\mathcal{M}}_{g,1}\left(X,d\right)\right]^{vir}}\frac{ev^{*}\left(\phi_{p}\right)}{\left(\alpha-\psi\right)}.\label{eq:delocalized-separation-of-disk}
\end{equation}

\section{The Gamma Class Formula\label{sec:Main-Results}}

\subsection{A formula for open Gromov-Witten invariants}

The proposed formula for open Gromov-Witten invariants is obtained
by composition of a disk function $\Delta_{X,L}$ and a descendant
invariant. $\Delta_{X,L}$ is built from combinatorial data about
the moment polytope, and a characteristic class. Recall that if $\delta_{i}$
are the Chern roots of a complex vector bundle $E$, Iritani's gamma
class \cite{Iritani} is a characteristic class associated to $E$
defined by 
\[
\Gamma_{E}:=\prod_{\delta_{i}}\Gamma\left(1+\delta_{i}\right).
\]
As observed in \cite{BCR-Gamma-fun}, in some cases the disk term
$D_{X,L}$ \eqref{eq:DXL} can be expressed using gamma classes. The
gamma class also appears in quantum cohomology, and can be regarded
as a localization contribution from constant maps in Floer theory
\cite{GammaClass}. The inputs of $\Delta_{X,L}$ are the torus weight
$\lambda$ of a normal direction to $f\left(\partial\Sigma\right)$,
and the homogenized Iritani gamma class $\hat{\Gamma}_{X}\in H^{*}\left(X;\mathbb{Q}\right)\left(z\right)$,
defined by 
\[
\widehat{\Gamma}_{X}:=\prod_{\delta_{i}}\Gamma\left(1+\frac{\delta_{i}}{z}\right),
\]
where $\delta_{i}$ are the Chern roots of the tangent bundle $T_{X}$.
(When $\deg z=2$, $\deg\widehat{\Gamma}_{X}=0$). With these definitions,
the main result is: 
\begin{thm}
\label{thm:main-result}Let $X$ be a Calabi-Yau 3-fold and $L\subset X$
a Lagrangian submanifold. Let $S^{1}$ act on $X$ such that the $S^{1}$
action preserves $L$, and $L$ intersects a rigid circle-invariant
curve $C$. Suppose that $L$ can be described in a neighborhood of
$L\cap C$ as the fixed locus of an anti-holomorphic involution. Let
$\gamma\in H^{2}\left(X;\mathbb{Q}\right)$. Then, the genus $g$,
1 boundary component, degree $d$, winding $w$ open Gromov-Witten
invariant with Lagrangian boundary $L$ is 
\begin{equation}
\left\langle \gamma\right\rangle _{d,w}^{g,1}=\left.\left(\Delta_{X,L}\circ\left\langle \gamma,\frac{\phi_{p}}{z-\psi}\right\rangle _{g,d}\right)\right|_{z=\alpha},\label{eq:open-GW-formula}
\end{equation}
where $\Delta_{X,L}$ is the disk function 
\begin{equation}
\Delta_{X,L}\left(\gamma\right):=\frac{\pi}{wz\widehat{\Gamma}_{X}\sin\left(\pi\frac{\lambda}{z}\right)}\cdot\gamma.\label{eq:disk-term-Delta}
\end{equation}
Here, $\widehat{\Gamma}_{X}$ is the homogeneous Iritani gamma class,
$\lambda$ is the weight of the $S^{1}$ action along a normal direction
to $C$, $\alpha=c_{1}(T_{0}\Delta)$ is the equivariant Chern class
of the induced representation of $S^{1}$ at the attachment point
of the disk, and $\phi_{p}$ is the equivariant class of the image
$p\in X$ of the disk attachment point. \end{thm}
\begin{rem*}
The setup in theorem~\ref{thm:main-result} is depicted in Figure~\ref{fig:local-vertex}.
In \cite{BCR-Gamma-fun}, the authors study the orbifold version of
this scenario with a Lagrangian obtained by the anti-holomorphic involution
$\sigma\left(\xi,x,y\right)=\left(1/\overline{\xi},\overline{\xi y},\overline{\xi x}\right)$,
and obtain an analogous result. The disk term \eqref{eq:disk-term-Delta}
only differs from \cite[equation 28]{BCR-Gamma-fun} in notation:
here, $z=\alpha$, and the winding $w$ is included to account for
automorphisms of the leg. 

Although the result above is stated in a more general form, it will
be shown below (lemma \ref{lem:involution-form}) that after a change
of coordinates, $\sigma\left(\xi,x,y\right)=\left(1/\overline{\xi},\overline{\xi y},\overline{\xi x}\right)$
is actually the only $S^{1}$-equivariant anti-holomorphic involution
satisfying these assumptions. In particular, the result proven in
this note applies to Aganagic-Vafa branes \cite{AV}. This author
is unsure how to prove the result without any assumptions on the local
geometry of $L$---in order to apply localization, one requires a
decription of the torus-fixed disks with boundary on $L$. However,
as will be shown in Section~\ref{sec:torus-knot-lagrangians}, this
result also applies without modification to a family of Lagrangian
cycles which are not obtained from the fixed locus of an anti-holomorphic
involution. 

\begin{figure}[th]
\begin{center} 
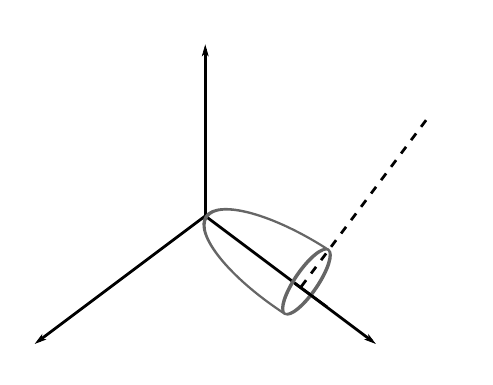
\end{center}

\caption{\label{fig:local-vertex}A local picture near a vertex in the toric
polytope.}

\begin{minipage}[t]{1\columnwidth}%
A Lagrangian $L$, obtained as the fixed locus of an anti-holomorphic
involution, intersects an edge of the toric polytope, labeled by the
local coordinate $\xi$. The only torus-fixed disks are the hemispheres
$D$ with $\partial D=L\cap\left\{ x=y=0\right\} \cong S^{1}$. Such
a map is given locally by $t\mapsto\left(\xi=t^{w},x=0,y=0\right)$.
The $x$-$y$ hyperplane is normal to the disk. The normal directions
to $C$ are spanned by $\partial_{x}$ and $\partial_{y}$, so the
weight $\lambda$ appearing in $\Delta_{X,L}$ \eqref{eq:disk-term-Delta}
can be the weight of any $S^{1}$-invariant line spanned by these
vectors (for example, $\lambda_{x}$ or $\lambda_{y}$). %
\end{minipage}
\end{figure}
 
\end{rem*}
In the genus $0$ case, the closed Gromov-Witten invariants in formula
\eqref{eq:delocalized-separation-of-disk} also appear as terms in
Givental's $J$ function \cite{Givental}. Givental's $J$ function
is the map on quantum cohomology $J_{X}\colon H^{*}\left(X;\mathbb{Q}\right)\longrightarrow H^{*}\left(X;\mathbb{Q}\right)\left(z\right)$
given by 
\[
J_{X}\left(\gamma\right)=z+\gamma+\sum_{n=0}^{\infty}\sum_{d\in H_{2}\left(X;\mathbb{Z}\right)}\left\langle \gamma^{n},\frac{T^{\alpha}}{z-\psi}\right\rangle _{0,d}T_{\alpha},
\]
where $T^{\alpha}$ is a basis for the cohomology of $X$, $T_{\alpha}$
is the dual basis with respect to the Poincar\'{e} pairing, and 
\[
\left\langle \gamma^{n},\frac{T^{\alpha}}{z-\psi}\right\rangle _{0,d}=\sum_{k=0}^{\infty}z^{-\left(k+1\right)}\left\langle \gamma^{n},\tau_{k}T^{\alpha}\right\rangle _{0,d}
\]
is a power series of gravitational descendant closed Gromov-Witten
invariants. To obtain the genus-$g$ generating function, a higher-genus
version of the $J$-function is needed. Define the genus-$g$ modified
$J$-function $J_{X}^{g}\colon H^{*}\left(X;\mathbb{Q}\right)\longrightarrow H^{*}\left(X;\mathbb{Q}\right)\left(z\right)$
to be 
\begin{equation}
J_{X}^{g}\left(\gamma\right)=z+\gamma+\sum_{n=0}^{\infty}\sum_{d\in H_{2}\left(X;\mathbb{Z}\right)}\frac{q^{d}}{n!}\left\langle \gamma^{n},\frac{T^{\alpha}}{z-\psi}\right\rangle _{g,d}T_{\alpha},\label{eq:modified-J-function}
\end{equation}
where 
\[
q^{d}=e^{2\pi i\int_{d}\omega}
\]
and $\omega$ is the complexified K\"{a}hler class of $X$. From
\eqref{eq:modified-J-function} and theorem~\ref{thm:main-result},
it is easy to write a generating function for open Gromov-Witten invariants.
The generating function for the one-boundary-component, winding $w$
open Gromov-Witten invariants $\left\langle \gamma\right\rangle {}_{d,w}^{g,1}$
is the function 
\[
\Phi_{w}\left(\gamma\right):=\sum_{g\geq0}\sum_{n\geq0}\sum_{d\in H_{2}\left(X;\mathbb{Z}\right)}g_{s}^{2g-1}\frac{q^{d}}{n!}\left\langle \gamma^{n}\right\rangle _{d,w}^{g,1},
\]
where $g_{s}$ is the string coupling constant, $\gamma\in H^{*}\left(X;\mathbb{Q}\right)$,
and the summation is only over combinations of $g$, $n$, and $d$
where the summands are defined. 
\begin{cor}
\label{cor:generating-function}Let $X$ and $L$ be as in Theorem~\ref{thm:main-result}.
Then, a generating function for the winding-$w$ open Gromov-Witten
invariants of $\overline{\mathcal{M}}_{g,1}\left(X,L;d,w\right)$
is given by the formula 
\[
\Phi_{w}\left(\gamma\right)=\sum_{g\geq0}g_{s}^{2g-1}\left.\left(\Delta_{X,L}\circ J_{X}^{g}\left(\gamma,\phi_{p}\right)\right)\right|_{z=\alpha.}
\]
\end{cor}
\begin{rem*}
Two observations about this generating function merit mention. First,
in \cite{BrCav,Brini-J-fun}, the authors use a similar procedure
to obtain a generating function for open invariants from a modification
of the $J$ function. The main distinction here is the presentation
of the disk term. Second, for $\gamma=1$, $\Phi_{w}$ has the expression
\[
\Phi_{w}=\sum_{g\geq0}\sum_{d\in H_{2}\left(X;\mathbb{Z}\right)}g_{s}^{2g-1}q^{d}GW_{d,w}^{g,1}.
\]
As will be discussed in Section~\ref{sec:torus-knot-lagrangians},
\cite{DSV} have found that, for a certain class of Lagrangian cycles
originating from torus knots, the expression above encodes the HOMFLY
polynomial associated to the original knot. 
\end{rem*}

\subsection{Proof of the main result}

The main content of the proof of theorem~\ref{thm:main-result} is
the comparison of the predicted disk term from \eqref{eq:disk-term-Delta}
with an explicit localization calculation of the open Gromov-Witten
invariant.

\subsubsection{Virtual localization of open Gromov-Witten invariants}

First, recall the virtual localization technique: As described in
Section~\ref{sub:Separating-the-disc}, the open Gromov-Witten invariant
$GW_{d,w}^{g}$ can be expressed as a product of a disk term $D_{X,L}$
and a descendant invariant. The surface $\Sigma$ can be written as
$\Sigma_{0}\cup_{\nu}\Delta$, with $\Sigma_{0}$ is a closed surface,
$\Delta$ a disk, and $\nu$ the point of attachment. In terms of
the cohomology of sheaves over $\Delta$, $D_{X,L}$ was found to
have the following expression: 
\[
D_{X,L}=\left(\frac{1}{w}\right)\frac{e_{S^{1}}\left(H^{1}\left(\Delta,T_{(\Delta,f_{\Delta})}\right)\right)}{e_{S^{1}}\left(H^{0}\left(\Delta,T_{(\Delta,f_{\Delta})}\right)\right)},\tag{\ref{eq:DXL}}
\]
where $f_{\Delta}:\Delta\rightarrow X$ is the restriction of the
map $f:\Sigma\rightarrow X$ to the disk $\Delta$, $p=f\left(\nu\right)$,
and $e_{S^{1}}\left(\cdot\right)$ denotes the $S^{1}$-equivariant
Euler classes of the specified bundles. To compute the disk contribution
to $GW_{d,w}^{g}$, one must compute each of these cohomology groups. 

In contrast to the analogous computation of closed invariants, the
Lagrangian $L$ imposes boundary conditions on the sections of $T_{\left(\Delta,f_{\Delta}\right)}$.
Let $f_{\partial}$ denote $f_{\Delta}|_{\partial\Delta}$. Then,
$T_{\left(\Delta,f_{\Delta}\right)}$ consists of sections of $f_{\Delta}^{*}T_{X}$
satisfying $s|_{\partial\Delta}\in f_{\partial}^{*}T_{L}$. To obtain
an explicit presentation of the boundary conditions, let $Ann\left(L\right)\subset T_{X}^{*}|_{L}$
be the subbundle of the cotangent bundle $T_{X}^{*}$ which annihilates
the tangent bundle $T_{L}\subset T_{X}|_{L}$. Choose a basis of sections
$\alpha_{1},\alpha_{2},\alpha_{3}$ of $Ann\left(L\right)$ along
the boundary $\partial D$ of the disk. (The $\alpha_{i}$ can be
obtained by, for example, linearizing the equations defining $L$).
$T_{\left(\Delta,f_{\Delta}\right)}$ consists of the sheaf of germs
of holomorphic sections of the bundle $f_{\Delta}^{*}T_{X}$ satisfying
the boundary conditions 
\begin{align}
f_{\partial}^{*}\left(\alpha_{j}\right)\left(s|_{\partial\Delta}\right) & =0, & j & =1,2,3.\label{eq:localization-bound-conds}
\end{align}
With this presentation of the boundary conditions, computing $H^{i}\left(\Delta,T_{\left(\Delta,f_{\Delta}\right)}\right)$
becomes an exercise in \v{C}ech cohomology: Let 
\begin{align*}
U & =\left\{ t:0<\left|t\right|\leq1\right\} , & U' & =\left\{ t:0\leq\left|t\right|<1\right\} 
\end{align*}
be an open cover of $\Delta$, and let $x,y,\xi$ be local coordinates
such that $f\left(\nu\right)=p$ is the origin $\left(x,y,\xi\right)=\left(0,0,0\right)$.
Then, local sections over $U$ and $U'$ are of the form 
\begin{eqnarray*}
s & = & \sum_{k\in\mathbb{Z}}\left(a_{k}t^{k}\partial_{x}+b_{k}t^{k}\partial_{y}+c_{k}t^{k}\partial_{\xi}\right),\\
s' & = & \sum_{k\geq0}\left(a_{k}'t^{k}\partial_{x}+b_{k}'t^{k}\partial_{y}+c_{k}'t^{k}\partial_{\xi}\right),
\end{eqnarray*}
and the coefficients $a_{k}$, $b_{k}$, and $c_{k}$ are subject
to boundary conditions imposed by \eqref{eq:localization-bound-conds}.
Finally, to apply localization, let $\rho_{\theta}:X\rightarrow X$
denote the $S^{1}$ action determined by 
\begin{align*}
\rho_{\theta}\left(x,y,\xi\right) & =\left(e^{i\lambda_{x}\theta}x,e^{i\lambda_{y}\theta}y,e^{i\lambda_{\xi}\theta}\xi\right), & \theta & \in S^{1}.
\end{align*}
The weights $\lambda_{x}$, $\lambda_{y}$, $\lambda_{\xi}$ are required
to satisfy $\lambda_{x}+\lambda_{y}+\lambda_{\xi}=0$ so that the
holomorphic volume form is preserved by the $S^{1}$ action.

\subsubsection{Boundary conditions}

Now, suppose that near the intersection $L\cap C$, $L$ is described
as the fixed locus of an anti-holomorphic involution $\sigma$. Choose
the local coordinates $x$, $y$, and $\xi$ such that $L\cap C$
is defined by $x=y=0$, $\left|\xi\right|^{2}=1$. 
\begin{lem}
\label{lem:involution-form}For a Calabi-Yau $S^{1}$ action on $\mathbb{C}^{3}$,
the only $S^{1}$-equivariant anti-holomorphic involution defined
on $\left(x,y,\xi\right)\in\mathbb{C}^{2}\times\mathbb{C}^{*}$ and
fixing the circle $x=y=0$, $\left|\xi\right|^{2}=1$ is 
\[
\sigma\left(x,y,\xi\right)=\left(X_{0}\overline{\xi y},Y_{0}\overline{\xi x},\frac{1}{\overline{\xi}}\right),
\]
where $X_{0},Y_{0}\in\mathbb{C}^{*}$ are two constants satisfying
$X_{0}\overline{Y_{0}}=1$. 
\end{lem}
Note that the Calabi-Yau condition means that the holomorphic 3-form
$dx\wedge dy\wedge d\xi$ is preserved by the $S^{1}$ action, i.e.,
if $S^{1}$ acts with weights $\left(\lambda_{x},\lambda_{y},\lambda_{\xi}\right)$,
then $\lambda_{x}+\lambda_{y}+\lambda_{\xi}=0$. 
\begin{proof}
$S^{1}$-equivariance is the requirement that $\rho_{\theta}\circ\sigma=\sigma\circ\rho_{\theta}$,
and is a necessary condition to apply localization to $L$. (By assumption,
in a neighborhood of $L\cap C$, $p\in L$ must satisfy $\rho_{\theta}\left(p\right)=\sigma\left(\rho_{\theta}\left(p\right)\right)$.
Because $L$ is defined locally by $p=\sigma\left(p\right)$, this
is equivalent to $\rho_{\theta}\circ\sigma=\sigma\circ\rho_{\theta}$).

Generically, $\sigma$ takes the form 
\[
\sigma\left(x,y,\xi\right)=\left(\sigma_{x},\sigma_{y},\sigma_{\xi}\right),
\]
where $\sigma_{x}$, $\sigma_{y}$, and $\sigma_{\xi}$ are Laurent
series in the variables $\overline{x}$, $\overline{y}$, $\overline{\xi}$:
\begin{align*}
\sigma_{x} & =\sum_{j,k,l\in\mathbb{Z}}X_{jkl}\overline{x}^{j}\overline{y}^{k}\overline{\xi}^{l},\\
\sigma_{y} & =\sum_{j,k,l\in\mathbb{Z}}Y_{jkl}\overline{x}^{j}\overline{y}^{k}\overline{\xi}^{l},\\
\sigma_{\xi} & =\sum_{j,k,l\in\mathbb{Z}}Z_{jkl}\overline{x}^{j}\overline{y}^{k}\overline{\xi}^{l}.
\end{align*}
$S^{1}$-equivariance imposes restrictions on the coefficients $X_{j,k,l}$,
$Y_{j,k,l}$, $Z_{j,k,l}$. For example, $\sigma_{\xi}$ must satisfy
\[
\sigma_{\xi}\left(e^{i\lambda_{x}\theta}x,e^{i\lambda_{y}\theta}y,e^{i\lambda_{\xi}\theta}\xi\right)=e^{i\lambda_{\xi}\theta}\sigma_{\xi}\left(x,y,\xi\right)
\]
for all $\theta\in S^{1}$. Expanding, this is 
\[
\sum_{j,k,l}Z_{j,k,l}e^{-i\theta\left(j\lambda_{x}+k\lambda_{y}+l\lambda_{\xi}\right)}\overline{x}^{j}\overline{y}^{k}\overline{\xi}^{l}=e^{i\lambda_{\xi}\theta}\sum_{j,k,l}Z_{j,k,l}\overline{x}^{j}\overline{y}^{k}\overline{\xi}^{l}.
\]
Hence, $-\left(l+1\right)\lambda_{\xi}=j\lambda_{x}+k\lambda_{y}$.
Recalling that $\lambda_{x}+\lambda_{y}=-\lambda_{\xi}$, this forces
$j=k=l+1$. Similar results hold for $\sigma_{x}$ and $\sigma_{y}$.
Therefore, anti-holomorphic involutions commuting with the $S^{1}$
action along $L$ must take the form 
\begin{align*}
\sigma_{x} & =\left(\overline{y\xi}\right)\sum_{l\in\mathbb{Z}}X_{l}\left(\overline{xy\xi}\right)^{l}, & \sigma_{y} & =\left(\overline{x\xi}\right)\sum_{l\in\mathbb{Z}}Y_{l}\left(\overline{xy\xi}\right)^{l}, & \sigma_{\xi} & =\left(\overline{xy}\right)\sum_{l\in\mathbb{Z}}Z_{l}\left(\overline{xy\xi}\right)^{l}.
\end{align*}
In fact there are further restrictions on $\sigma$. Because $\sigma$
must be defined along $\left\{ x=y=0,\left|\xi\right|^{2}=1\right\} $,
$X_{l}=Y_{l}=0$ for $l<0$, and $Z_{-1}=1$, $Z_{l}=0$ for $l<-1$.
Substituting these relations into $\sigma^{2}=1$, the equation $x=\sigma_{x}\circ\sigma\left(x,y,\xi\right)$
becomes 
\begin{eqnarray*}
x & = & \left(x\sum_{l\geq0}\overline{Y_{l}}\left(xy\xi\right)^{l}+x^{2}y\xi\sum_{j,k\geq0}\overline{Y_{j}Z_{k}}\left(xy\xi\right)^{j+k}\right)\\
 &  & \times\sum_{l\geq0}X_{l}\left(\frac{1}{\xi}\sum_{j,k\geq0}\overline{X_{j}Y_{k}}\left(xy\xi\right)^{j+k+1}+\sum_{i,j,k\geq0}\overline{X_{i}Y_{j}Z_{k}}\left(xy\xi\right)^{i+j+k+2}\right)^{l}\\
 & = & x\overline{Y_{0}}X_{0}+\left(x^{2}y\xi\right)g\left(x,y,\xi\right),
\end{eqnarray*}
where $g$ is a power series in $x,y,\xi$. Therefore, $X_{0}\overline{Y_{0}}=1$,
and, for degree reasons, $g\left(x,y,\xi\right)=0$. An analogous
relation holds for $\sigma_{y}$. Similarly, $\xi=\sigma_{\xi}\circ\sigma\left(x,y,\xi\right)$
becomes 
\begin{eqnarray*}
\xi & = & \xi-\sum_{k\geq2}\left(-1\right)^{k}\xi^{k}\left(xy\sum_{l\geq0}\overline{Z_{l}}\left(xy\xi\right)^{l}\right)^{k}\\
 &  & +\left(\xi^{2}xy\sum_{j,k\geq0}\overline{X_{j}Y_{k}}\left(xy\xi\right)^{j+k}\right)\\
 &  & \times\sum_{l\geq0}Z_{l}\left(\frac{1}{\xi}\sum_{j,k\geq0}\overline{X_{j}Y_{k}}\left(xy\xi\right)^{j+k+1}+\sum_{i,j,k\geq0}\overline{X_{i}Y_{j}Z_{k}}\left(xy\xi\right)^{i+j+k+2}\right)^{l}\\
 & = & \xi+\left(\xi^{2}xy\right)h\left(x,y,\xi\right).
\end{eqnarray*}
As with $\sigma_{x}$, $h\left(x,y,\xi\right)=0$. A careful examination
of the coefficients in the above equations reveals that the only solutions
to $\sigma^{2}=1$ are $Z_{l}=0$ for $l\geq0$ and $X_{l}=Y_{l}=0$
for $l\geq1$. This completes the proof of the lemma. 
\end{proof}
Applying the above lemma, the defining equations of $L$ near $L\cap C$
are 
\begin{align*}
x & =X_{0}\overline{y\xi}, & y & =Y_{0}\overline{x\xi}, & \xi & =\frac{1}{\overline{\xi}}.
\end{align*}
Linearizing these equations yields 
\begin{align*}
dx & =X_{0}\left(\overline{\xi}d\overline{y}+\overline{y}d\overline{\xi}\right), & dy & =Y_{0}\left(\overline{\xi}d\overline{x}+\overline{x}d\overline{\xi}\right), & d\xi & =-\left(\frac{1}{\overline{\xi}}\right)^{2}d\overline{\xi}^{2},
\end{align*}
and at $x=y=0$, these equations simplify to 
\begin{align*}
dx & =X_{0}\overline{\xi}d\overline{y}, & dy & =Y_{0}\overline{\xi}d\overline{x}, & d\xi & =-\left(\frac{1}{\overline{\xi}}\right)^{2}d\overline{\xi}.
\end{align*}
As $\xi=\overline{\xi}^{-1}$ along $\left|\xi\right|^{2}=1$, a basis
for $Ann\left(L\right)$ along $\partial D$ is given by the 1-forms
\begin{align*}
\alpha_{1} & =dx-X_{0}\overline{\xi}d\overline{y}, & \alpha_{2} & =dy-Y_{0}\overline{\xi}d\overline{x}, & \alpha_{3} & =\overline{\xi}d\xi+\xi d\overline{\xi}.
\end{align*}

Recall that local sections over $U=\left\{ t:0<\left|t\right|\leq1\right\} $
are of the form 
\[
s=\sum_{k\in\mathbb{Z}}\left(a_{k}t^{k}\partial_{x}+b_{k}t^{k}\partial_{y}+c_{k}t^{k}\partial_{\xi}\right).
\]
Parameterizing $\left|t\right|=1$ by $e^{i\theta}=t$, the map $f_{\partial}$
takes the form $f_{\partial}\left(e^{i\theta}\right)=\left(x=0,y=0,\xi=e^{iw\theta}\right)$.
The boundary conditions $f_{\partial}^{*}\left(\alpha_{j}\right)\left(s|_{\partial\Delta}\right)=0$
impose restrictions on the coefficients $a_{k}$, $b_{k}$, $c_{k}$:
\begin{eqnarray*}
f_{\partial}^{*}\left(\alpha_{1}\right)\left(s|_{\partial\Delta}\right) & = & \left(dx-X_{0}e^{-iw\theta}d\overline{y}\right)\left(\sum_{k\in\mathbb{Z}}\left(a_{k}e^{ik\theta}\partial_{x}+b_{k}e^{ik\theta}\partial_{y}+c_{k}e^{ik\theta}\partial_{\xi}\right)\right)\\
 & = & \sum_{k\in\mathbb{Z}}\left(a_{k}e^{ik\theta}-X_{0}e^{-iw\theta}\overline{b_{k}}e^{-ik\theta}\right),\\
f_{\partial}^{*}\left(\alpha_{2}\right)\left(s|_{\partial\Delta}\right) & = & \left(dy-Y_{0}e^{-iw\theta}d\overline{x}\right)\left(\sum_{k\in\mathbb{Z}}\left(a_{k}e^{ik\theta}\partial_{x}+b_{k}e^{ik\theta}\partial_{y}+c_{k}e^{ik\theta}\partial_{\xi}\right)\right)\\
 & = & \sum_{k\in\mathbb{Z}}\left(b_{k}e^{ik\theta}-Y_{0}e^{-iw\theta}\overline{a}_{k}e^{-ik\theta}\right),\\
f_{\partial}^{*}\left(\alpha_{3}\right)\left(s|_{\partial\Delta}\right) & = & \left(e^{-iw\theta}d\xi+e^{iw\theta}d\overline{\xi}\right)\left(\sum_{k\in\mathbb{Z}}\left(a_{k}e^{ik\theta}\partial_{x}+b_{k}e^{ik\theta}\partial_{y}+c_{k}e^{ik\theta}\partial_{\xi}\right)\right)\\
 & = & \sum_{k\in\mathbb{Z}}\left(c_{k}e^{i\left(k-w\right)\theta}+\overline{c}_{k}e^{-i\left(k-w\right)\theta}\right).
\end{eqnarray*}
These yield the following equations for the coefficients $a_{k}$,
$b_{k}$, $c_{k}$: 
\begin{align*}
a_{k}-X_{0}\overline{b}_{-k-w} & =0, & b_{k}-Y_{0}\overline{a}_{-k-w} & =0, & c_{k}+\overline{c}_{2w-k} & =0.
\end{align*}
The first two equations are actually equivalent: after complex conjugation,
relabeling of indices, and substituting $X_{0}\overline{Y_{0}}=1$,
the second equation becomes the first. So, the Lagrangian boundary
conditions on sections over $U$ are 
\begin{align}
a_{k} & =X_{0}\overline{b}_{-k-w}, & c_{k} & =\overline{c}_{2w-k}.\label{eq:Cech-lag-boundary-cond}
\end{align}
From these boundary conditions, the cohomology groups $H^{i}\left(\Delta,T_{\left(\Delta,f_{\Delta}\right)}\right)$
can be computed explicitly.

\subsubsection{Computation of cohomology groups and equivariant classes}

$H^{0}\left(\Delta,T_{\left(\Delta,f_{\Delta}\right)}\right)$ consists
of the global sections, i.e., holomorphic sections $s$ on $\Delta$.
These take the form 
\[
s=\sum_{k\geq0}\left(a_{k}t^{k}\partial_{x}+b_{k}t^{k}\partial_{y}+c_{k}t^{k}\partial_{\xi}\right),
\]
with $a_{k}$, $b_{k}$ and $c_{k}$ subject to the boundary conditions
in \eqref{eq:Cech-lag-boundary-cond}, and $a_{k}=b_{k}=c_{k}=0$
for $k<0$. In particular, the equation $a_{k}=X_{0}\overline{b}_{-k-w}$
implies that $a_{k}=0$ for all $k$. Shifting indices $k\rightarrow-k-w$,
this equation also implies that $b_{k}=0$ for all $k$. Finally,
from the last boundary equation $c_{k}=\overline{c}_{2w-k}$, $c_{k}=0$
for $k>2w$. So, $H^{0}\left(\Delta,T_{\left(\Delta,f_{\Delta}\right)}\right)$
consists of sections of the form 
\[
s=\sum_{k=0}^{w-1}\left(c_{k}t^{k}\partial_{\xi}+\overline{c}_{k}t^{2w-k}\partial_{\xi}\right)+c_{w}t^{w}\partial_{\xi},
\]
where $c_{w}$ is real. As a vector space, $H^{0}\left(\Delta,T_{\left(\Delta,f_{\Delta}\right)}\right)$
is isomorphic to 
\[
\mathbb{R}\left\langle t^{w}\partial_{\xi}\right\rangle \oplus\bigoplus_{k=0}^{w-1}\mathbb{C}\left\langle t^{k}\partial_{\xi}\right\rangle .
\]
The map $f_{\Delta}$ takes $t\mapsto\left(x=0,y=0,\xi=t^{w}\right)$,
so $S^{1}$ action the section $t^{k}\partial_{\xi}$ with weight
$\lambda_{\xi}\left(k/w-1\right)$. $t^{w}\partial_{\xi}$ is fixed
by the $S^{1}$ action, so $H^{0}\left(\Delta,T_{\left(\Delta,f_{\Delta}\right)}\right)^{m}$
is just the complex part of this vector space. Hence, 
\begin{equation}
e_{S^{1}}\left(H^{0}\left(\Delta,T_{\left(\Delta,f_{\Delta}\right)}\right)\right)=\prod_{k=0}^{w-1}\lambda_{\xi}\left(\frac{k}{w}-1\right).\label{eq:main-thm-pf-h0}
\end{equation}

$H^{1}\left(\Delta,T_{\left(\Delta,f_{\Delta}\right)}\right)$ consists
of the cokernel to the \v{C}ech differential. Sections over $U\cap U'$
can be written as 
\begin{align*}
\delta & =\sum_{k\in\mathbb{Z}}\left(\alpha_{k}t^{k}\partial_{x}+\beta_{k}t^{k}\partial_{y}+\gamma_{k}t^{k}\partial_{\xi}\right)\\
 & =\sum_{k\leq-w}\alpha_{k}t^{k}\partial_{x}+\sum_{k=1-w}^{-1}\alpha_{k}t^{k}\partial_{x}+\sum_{k\geq0}\alpha_{k}t^{k}\partial_{x}\\
 & +\sum_{k<0}\beta_{k}t^{k}\partial_{y}+\sum_{k\geq0}\beta_{k}t^{k}\partial_{y}\\
 & +\sum_{k<0}\gamma_{k}t^{k}\partial_{\xi}+\sum_{k\geq0}\gamma_{k}t^{k}\partial_{\xi}.
\end{align*}
The image of the \v{C}ech differential consists of sections $\delta$
of the form $\delta=s-s'$. In terms of the coefficients, this is
\begin{align*}
\alpha_{k} & =\begin{cases}
a_{k}, & k<0\\
a_{k}-a_{k}', & k\geq0
\end{cases}, & \beta_{k} & =\begin{cases}
b_{k}, & k<0\\
b_{k}-b_{k}', & k\geq0
\end{cases}, & \gamma_{k} & =\begin{cases}
c_{k}, & k<0\\
c_{k}-c_{k}', & k\geq0,
\end{cases}
\end{align*}
where again, $a_{k}$, $b_{k}$, and $c_{k}$ are subject to the boundary
conditions \eqref{eq:Cech-lag-boundary-cond}. Solutions always exist
for $\gamma_{k}$: set $c_{k}=\gamma_{k}$ for $k<0$, and set $c_{k}'=\overline{c}_{2w-k}-\gamma_{k}$
for $k\geq0$. Similarly, because $a_{k}'$ and $c_{k}'$ are completely
free, any $\alpha_{k}$ and $\beta_{k}$ for $k\geq0$ can be solved
for. However, to solve for $\alpha_{k}$ and $\beta_{k}$ for $k<0$,
it must be the case that $b_{k}=\beta_{k}$. The first boundary equation
$a_{k}=X_{0}\overline{b}_{-k-w}$ then implies that $a_{k}$ is fixed
for $-w<k<0$. So, there are no solutions if $\alpha_{k}\neq X_{0}\beta_{-k-w}$
in $-w<k<0$. When $k\leq-w$, $-k-w\geq0$, so setting $b_{-k-w}=\alpha_{k}$
and $b_{-k-w}'=\alpha_{k}-\beta_{-k-w}$ will solve these equations.
Therefore, the cokernel of the \v{C}ech differential is isomorphic
to the space of sections $\delta$ of the form 
\[
\delta=\sum_{k=1-w}^{-1}\alpha_{k}t^{k}\partial_{x}.
\]
The induced $S^{1}$ action on $t^{k}\partial_{x}$ has weight $\frac{k}{w}\lambda_{\xi}-\lambda_{x}$.
As a vector space, $H^{1}\left(\Delta,T_{\left(\Delta,f_{\Delta}\right)}\right)$
is 
\[
\bigoplus_{k=1-w}^{-1}\mathbb{C}\left\langle t^{k}\partial_{x}\right\rangle ,
\]
and 
\begin{equation}
e_{S^{1}}\left(H^{1}\left(\Delta,T_{\left(\Delta,f_{\Delta}\right)}\right)\right)=\prod_{k=1-w}^{-1}\left(\frac{k}{w}\lambda_{\xi}-\lambda_{x}\right).\label{eq:main-thm-pf-h1}
\end{equation}

\subsubsection{Comparison of disk terms\label{sub:Pf-comparing-disk-terms}}

Substituting \eqref{eq:main-thm-pf-h0} and \eqref{eq:main-thm-pf-h1}
in \eqref{eq:DXL} yields 
\begin{equation}
D_{X,L}=\frac{1}{w}\frac{\prod_{k=1-w}^{-1}\left(\frac{k}{w}\lambda_{\xi}-\lambda_{x}\right)}{\prod_{k=0}^{w-1}\lambda_{\xi}\left(\frac{k}{w}-1\right)}.\label{eq:disk-pf-thm}
\end{equation}
The proof will be complete if \eqref{eq:disk-pf-thm} is equivalent
to the claimed expression \eqref{eq:disk-term-Delta}: 
\[
\frac{1}{w}\frac{\prod_{k=1-w}^{-1}\left(\frac{k}{w}\lambda_{\xi}-\lambda_{x}\right)}{\prod_{k=0}^{w-1}\lambda_{\xi}\left(\frac{k}{w}-1\right)}=\left.\frac{\pi}{wz\widehat{\Gamma}_{X}\sin\left(\pi\frac{\lambda}{z}\right)}\right|_{z=\alpha.}
\]
First, observe that 
\[
\prod_{k=0}^{w-1}\lambda_{\xi}\left(\frac{k}{w}-1\right)=\left(-\frac{\lambda_{\xi}}{w}\right)^{w}\Gamma\left(w+1\right).
\]
Similarly, 
\begin{eqnarray*}
\prod_{k=1-w}^{-1}\left(\frac{k}{w}\lambda_{\xi}-\lambda_{x}\right) & = & \left(-\frac{\lambda_{\xi}}{w}\right)^{w-1}\prod_{k=1}^{w-1}\left(k+w\frac{\lambda_{x}}{\lambda_{\xi}}\right)\\
 & = & \left(-\frac{\lambda_{\xi}}{w}\right)^{w-1}\frac{\Gamma\left(w\frac{\lambda_{x}}{\lambda_{\xi}}+w\right)}{\Gamma\left(w\frac{\lambda_{x}}{\lambda_{\xi}}+1\right)}.
\end{eqnarray*}
Recall that, by assumption, $\lambda_{x}+\lambda_{y}+\lambda_{\xi}=0$,
so $\frac{\lambda_{x}}{\lambda_{\xi}}=-1-\frac{\lambda_{y}}{\lambda_{\xi}}$.
Therefore, $\Gamma\left(w\frac{\lambda_{x}}{\lambda_{\xi}}+w\right)=\Gamma\left(-w\frac{\lambda_{y}}{\lambda_{\xi}}\right)$.
The above manipulations show that 
\[
\frac{1}{w}\frac{\prod_{k=1-w}^{-1}\left(\frac{k}{w}\lambda_{\xi}-\lambda_{x}\right)}{\prod_{k=0}^{w-1}\lambda_{\xi}\left(\frac{k}{w}-1\right)}=\left(\frac{1}{w}\right)\frac{\Gamma\left(-w\frac{\lambda_{y}}{\lambda_{\xi}}\right)}{\left(-\frac{\lambda_{\xi}}{w}\right)\Gamma\left(w+1\right)\Gamma\left(w\frac{\lambda_{x}}{\lambda_{\xi}}+1\right)}.
\]
The induced action on $T_{0}\Delta$ carries weight $\alpha=\frac{\lambda_{\xi}}{w}$.
Substitute $\frac{\lambda_{\xi}}{w}=z$ and apply Euler's reflection
formula to get 
\begin{eqnarray*}
D_{X,L} & = & \left(-\frac{1}{w}\right)\frac{\Gamma\left(-\frac{\lambda_{y}}{z}\right)}{z\Gamma\left(\frac{\lambda_{\xi}}{z}+1\right)\Gamma\left(\frac{\lambda_{x}}{z}+1\right)}\\
 & = & \left(\frac{1}{w}\right)\frac{\pi}{z\Gamma\left(\frac{\lambda_{\xi}}{z}+1\right)\Gamma\left(\frac{\lambda_{x}}{z}+1\right)\Gamma\left(\frac{\lambda_{y}}{z}+1\right)\sin\left(\pi\frac{\lambda_{y}}{z}\right)}\\
 & = & \frac{\pi}{wz\widehat{\Gamma}_{X}\sin\left(\pi\frac{\lambda_{y}}{z}\right)}.
\end{eqnarray*}
Generically, there are two $S^{1}$-invariant normal directions to
$L\cap C$ in $X$, given by the tangent vectors $\partial_{x}$ and
$\partial_{y}$. In the formula in \eqref{eq:disk-term-Delta}, $\lambda$
may be the weight of either of these directions, i.e., $\lambda=\lambda_{y}$
or $\lambda_{x}$. The choice of $\lambda$ changes the sign of \eqref{eq:disk-term-Delta}
because $\sin\left(\pi\frac{\lambda_{y}}{z}\right)=-\sin\left(\pi\frac{\lambda_{x}}{z}\right)$.
This sign ambiguity reflects an overall choice of orientation of $\overline{\mathcal{M}}_{g,1,0}\left(X,L;d,w\right)$
\cite{AKV,GraberZaslow,KatzLiu}. This completes the proof of the
main result.

\section{Comparison to Known Localization Calculations\label{sec:Comparisons}}

In this section, \eqref{eq:open-GW-formula} is compared to previous
virtual localization calculations of open Gromov-Witten invariants.
In the interest of brevity, only the geometric setup and final results
are stated below; the reader interested in further localization calculations
is referred to the original sources, or the computation appearing
in the proof of theorem~\ref{thm:main-result}.

\subsection{Simple Lagrangians for $\mathcal{O}_{\mathbb{P}^{1}}\left(-1,-1\right)$\label{sub:KL-lagrangians}}

This situation was first described in \cite{KatzLiu}, and the gamma
class formula for an orbifold generalization of this case was given
in \cite{BCR-Gamma-fun}. For completeness, the scenario will also
be described here. Let $X=\mathcal{O}_{\mathbb{P}^{1}}\left(-1,-1\right)$.
$X$ appears often in Gromov-Witten theory and mirror symmetry: it
is the small resolution of the conifold singularity, the normal bundle
to a smooth rational line in a Calabi-Yau 3-fold, and it can be obtained
from a $U(1)$ gauge theory with 4 chiral fields with charges $\left(1,1,-1,-1\right)$.
$X$ can be described symplectically using symplectic reduction on
$\mathbb{C}^{4}$, and in this setting it is easiest to obtain the
moment polytope of $X$. 

Let $S^{1}$ act on $\mathbb{C}^{4}$ with weights $\left(1,1,-1,-1\right)$.
Then, the moment map for this action is 
\begin{eqnarray*}
\mu\colon\mathbb{C}^{4} & \longrightarrow & \mathfrak{s}^{1}\cong\mathbb{R}\\
\left(z_{1},z_{2},z_{3},z_{4}\right) & \mapsto & \frac{1}{2}\left(\left|z_{1}\right|^{2}+\left|z_{2}\right|^{2}-\left|z_{3}\right|^{2}-\left|z_{4}\right|^{2}\right),
\end{eqnarray*}
and it can be seen (for example, by choosing appropriate local coordinates
and checking transition functions) that 
\[
X\cong\mu^{-1}\left(\frac{r}{2}\right)/S^{1}=\left\{ \left|z_{1}\right|^{2}+\left|z_{2}\right|^{2}-\left|z_{3}\right|^{2}-\left|z_{4}\right|^{2}=r\right\} /S^{1}
\]
for $r\in\mathbb{R}_{>0}$ ($r$ determines the symplectic volume
of the base $\mathbb{P}^{1}$). There's a natural anti-holomorphic
involution $\sigma$ on $\mathbb{C}^{4}$ given by 
\[
\sigma\left(z_{1},z_{2},z_{3},z_{4}\right)=\left(\overline{z_{2}},\overline{z_{1}},\overline{z_{4}},\overline{z_{3}}\right).
\]
The fixed locus of this involution is a Lagrangian submanifold $\widetilde{L}$
of $\mathbb{C}^{4}$ defined by the equations 
\begin{align*}
\left|z_{1}\right|^{2} & =\left|z_{2}\right|^{2},\\
\left|z_{3}\right|^{2} & =\left|z_{4}\right|^{2},\\
\overline{z_{1}z_{2}z_{3}z_{4}} & =z_{1}z_{2}z_{3}z_{4}.
\end{align*}
 Because $\widetilde{L}$ is preserved by the $S^{1}$ action, $\mu^{-1}\left(\frac{r}{2}\right)\cap\widetilde{L}/S^{1}$
defines a Lagrangian $L\subset X$. 

\begin{figure}[th]
\begin{center} 
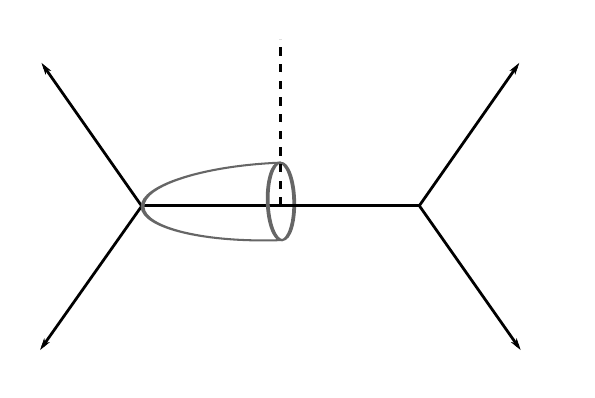
\end{center}

\caption{\label{fig:KL-lagrangian}The toric polytope for $X=\mathcal{O}_{\mathbb{P}^{1}}\left(-1,-1\right)$
with a Lagrangian.}

\begin{minipage}[t]{1\columnwidth}%
This figure depicts the geometry described in Section~\ref{sub:KL-lagrangians}.
A Lagrangian $L$, obtained as the fixed locus of an anti-holomorphic
involution, intersects the zero section of $X$. Local coordinates
$\left(x,y\right)$ and $\left(z,w\right)$ parametrize the fibers
of $X$ in a neighborhood of the two vertices of the moment polytope.
One possible torus-fixed disk $D$ with boundary $\partial D=L\cap\mathbb{P}^{1}\cong S^{1}$
is depicted.%
\end{minipage}
\end{figure}

This Lagrangian is easy to visualize in the moment polytope of $X$.
The moment polytope is the image of $X$ in $\mathbb{R}^{4}$ under
the projection $z_{i}\mapsto\left|z_{i}\right|^{2}$. Then, $L$ is
the intersection of the two planes $\left|z_{1}\right|^{2}=\left|z_{2}\right|^{2}$
and $\left|z_{3}\right|^{2}=\left|z_{4}\right|^{2}$ in the polytope.
$L$ intersects the zero section $\mathbb{P}^{1}$ along its equator,
so that 
\[
L\cap\mathbb{P}^{1}=\left\{ \left|z_{1}\right|^{2}=\left|z_{2}\right|^{2},\left|z_{3}\right|^{2}=\left|z_{4}\right|^{2}=0,\left|z_{1}\right|^{2}+\left|z_{2}\right|^{2}=r\right\} \cong S^{1},
\]
as depicted in Figure~\ref{fig:KL-lagrangian}. 
 There are two unique disks with boundary on $L$, corresponding to
the two hemispheres of the $\mathbb{P}^{1}$. In local coordinates
$\left(\xi,x,y\right)$ defined by 
\begin{align*}
\xi & =\frac{z_{1}}{z_{2}}, & x & =z_{2}z_{3}, & y & =z_{2}z_{4},
\end{align*}
$L$ is defined by the fixed locus of the anti-holomorphic involution
\[
\sigma\left(\xi,x,y\right)=\left(\frac{1}{\overline{\xi}},\overline{\xi y},\overline{\xi x}\right).
\]
It is readily checked that at $L\cap\mathbb{P}^{1}=\left\{ \left|\xi\right|^{2}=1\right\} $,
and the disk based at the $z_{1}=0$ pole of the $\mathbb{P}^{1}$
takes the form $\left|\xi\right|\leq1$. The winding $w$ disk map
is 
\[
t\mapsto\left(\xi=t^{w},x=0,y=0\right).
\]
In this situation, \cite{KatzLiu} computed 
\[
\left(\frac{1}{w}\right)\frac{e_{S^{1}}\left(H^{1}\left(\Delta,T_{(\Delta,f_{\Delta})}\right)\right)}{e_{S^{1}}\left(H^{0}\left(\Delta,T_{(\Delta,f_{\Delta})}\right)\right)}=\left(\frac{1}{w}\right)\frac{\prod_{k=1-w}^{-1}\left(\frac{k}{w}\lambda_{\xi}-\lambda_{x}\right)}{\prod_{k=0}^{w-1}\lambda_{\xi}\left(\frac{k}{w}-1\right)}.
\]
Comparing with \eqref{eq:disk-pf-thm} in Section~\ref{sub:Pf-comparing-disk-terms},
this is $i^{*}\left(\Delta_{X,L}\right)_{z=\alpha}$, so this result
agrees with the proposed formula.

\subsection{The canonical bundle of $\mathbb{P}^{2}$\label{sub:Kp2}}

\begin{figure}[th]
\begin{center} 
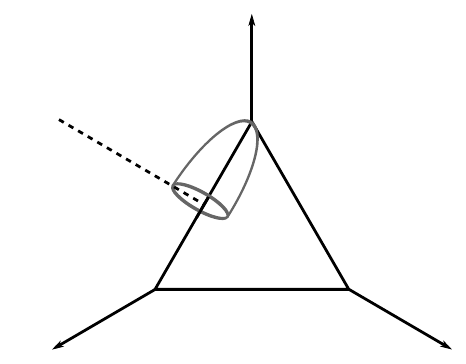
\end{center}

\caption{\label{fig:GZ-Lagrangian}The toric polytope for $X=\mathcal{O}_{\mathbb{P}^{2}}\left(-3\right)$
with a Lagrangian.}

\begin{minipage}[t]{1\columnwidth}%
This figure depicts the geometry described in Section~\ref{sub:Kp2}.
A Lagrangian $L$, obtained as the fixed locus of an anti-holomorphic
involution, intersects the zero section of $X$ along one edge of
the moment polytope. One possible torus-fixed disk $D$ with boundary
$\partial D=L\cap\mathbb{P}^{2}$ is depicted. %
\end{minipage}
\end{figure}

This situation was described in \cite{GraberZaslow}. As in the previous
example, $X=\mathcal{O}_{\mathbb{P}^{2}}\left(-3\right)$ can be obtained
via symplectic reduction. Let $S^{1}$ act on $\mathbb{C}^{4}$ with
weights $\left(1,1,1,-3\right)$. Then, 
\[
X\cong\mu^{-1}\left(\frac{r}{2}\right)/S^{1}=\left\{ \left|z_{1}\right|^{2}+\left|z_{2}\right|^{2}+\left|z_{3}\right|^{2}-3\left|z_{4}\right|^{2}=r\right\} /S^{1}
\]
for $r\in\mathbb{R}_{>0}$. \cite{GraberZaslow} consider the Lagrangian
submanifold $\widetilde{L}_{c}\subset\mathbb{C}^{4}$ defined by 
\begin{eqnarray*}
\left|z_{1}\right|^{2}-\left|z_{3}\right|^{2} & = & c,\\
\left|z_{2}\right|^{2}-\left|z_{4}\right|^{2} & = & 0,\\
z_{1}z_{2}z_{3}z_{4} & = & \overline{z_{1}z_{2}z_{3}z_{4}},
\end{eqnarray*}
where $-r<c<r$. Because $\widetilde{L}_{c}$ is preserved by the
$S^{1}$ action, it descends to a Lagrangian $L_{c}\subset X$, as
depicted in Figure~\ref{fig:GZ-Lagrangian}. 
 Note that $c$ parametrizes the intersection of $L_{c}$ with the
$\mathbb{P}^{1}$ given by the image of $\left|z_{1}\right|^{2}+\left|z_{3}\right|^{2}=r$
in the quotient space. At $c=0$, $L_{c}$ intersects this curve along
its equator. For simplicity, restrict attention to $L=L_{0}$ (locally,
other values of $c$ can be obtained by a coordinate transformation).
In local coordinates 
\begin{align}
\xi & =\frac{z_{1}}{z_{3}}, & x & =\frac{z_{2}}{z_{3}}, & y & =z_{3}^{3}z_{4},\label{eq:kp2-local-coords}
\end{align}
the Lagrangian $L$ is the fixed locus of the anti-holomorphic involution
\[
\sigma\left(\xi,x,y\right)=\left(\frac{1}{\overline{\xi}},\overline{\xi y},\overline{\xi x}\right).
\]
The disk is $\left|\xi\right|^{2}\leq1$, and the winding $w$ disk
map is 
\[
t\mapsto\left(\xi=t^{w},x=0,y=0\right).
\]
So, locally, the situation computed in \cite{GraberZaslow} is identical
to \cite{KatzLiu}. As seen in Section~\ref{sub:KL-lagrangians},
this agrees with theorem~\ref{thm:main-result}. 

\begin{figure}[th]
\begin{center} 
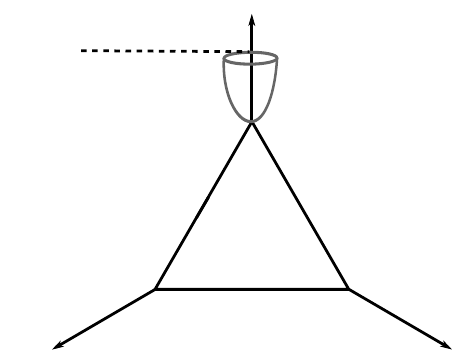
\end{center}

\caption{\label{fig:Kp2-Fiber-Lagrangian}$X=\mathcal{O}_{\mathbb{P}^{2}}\left(-3\right)$
with a Lagrangian brane on an external leg.}

\begin{minipage}[t]{1\columnwidth}%
A Lagrangian cycle intersects an external leg of the moment polytope
of $X$. The Lagrangian is obtained as the fixed locus of an anti-holomorphic
involution. There is only one torus-fixed disk $D$ with boundary
on $L$, as depicted above. %
\end{minipage}
\end{figure}

Slightly extending the computation in \cite{GraberZaslow}, theorem~\ref{thm:main-result}
can also be used to compute the invariants associated to a Lagrangian
cycle intersecting an external leg of the moment polytope. Let $\tilde{L}$
be the submanifold of $\mathbb{C}^{4}$ defined by 
\begin{eqnarray*}
\left|z_{1}\right|^{2}-\left|z_{2}\right|^{2} & = & 0,\\
\left|z_{3}\right|^{2}-\left|z_{4}\right|^{2} & = & 0,\\
z_{1}z_{2}z_{3}z_{4} & = & \overline{z_{1}z_{2}z_{3}z_{4}}.
\end{eqnarray*}
Again, these equations are preserved by the $S^{1}$ action, so the
image of $\tilde{L}$ in the quotient space $X$ is a well-defined
Lagrangian submanifold $L$. $L$ can be equivalently described as
the fixed locus of the anti-holomorphic involution $\left(z_{1},z_{2},z_{3},z_{4}\right)\mapsto\left(\overline{z_{2}},\overline{z_{1}},\overline{z_{4}},\overline{z_{3}}\right)$.
In local coordinates $\left(\xi,x,y\right)$ \eqref{eq:kp2-local-coords},
$L$ is defined by 
\begin{eqnarray*}
\left|y\right|^{2} & = & 1,\\
\left|x\right|^{2} & = & \left|\xi\right|^{2},\\
\xi xy & = & \overline{\xi xy}.
\end{eqnarray*}
(This Lagrangian cycle is shown in Figure~\ref{fig:Kp2-Fiber-Lagrangian}).
 The disk is $\left|y\right|^{2}\leq1$, and the winding $w$ disk
map is $t\mapsto\left(\xi=0,x=0,y=t^{w}\right)$. So, $\alpha=\lambda_{y}/w$.
Applying theorem~\ref{thm:main-result}, 
\begin{eqnarray*}
\left(\frac{1}{w}\right)\frac{e_{S^{1}}\left(H^{1}\left(\Delta,T_{(\Delta,f_{\Delta})}\right)\right)}{e_{S^{1}}\left(H^{0}\left(\Delta,T_{(\Delta,f_{\Delta})}\right)\right)} & = & \left.\frac{\pi}{wz\widehat{\Gamma}_{X}\sin\left(\pi\frac{\lambda_{\xi}}{z}\right)}\right|_{z=\alpha}\\
 & = & \frac{-\pi\Gamma\left(-w\frac{\lambda_{\xi}}{\lambda_{y}}\right)}{\lambda_{y}\Gamma\left(w\frac{\lambda_{x}}{\lambda_{y}}+1\right)\Gamma\left(w+1\right)}\\
 & = & \frac{1}{w}\frac{\prod_{k=1-w}^{-1}\left(\frac{k}{w}\lambda_{y}-\lambda_{\xi}\right)}{\prod_{k=0}^{w-1}\lambda_{y}\left(\frac{k}{w}-1\right)}.
\end{eqnarray*}
Here, the normal direction weight $\lambda=\lambda_{\xi}$ has been
chosen in \eqref{eq:disk-term-Delta}. Choosing $\lambda=\lambda_{x}$
instead changes the sign, reflecting the overall dependence of these
counts on the choice of torus weights. This can be seen from the product
identity: 
\[
\prod_{k=1-w}^{-1}\left(\frac{k}{w}\lambda_{y}-\lambda_{\xi}\right)=\left(-1\right)^{w-1}\prod_{k=1-w}^{-1}\left(\frac{k}{w}\lambda_{y}-\lambda_{x}\right).
\]

\section{Disk invariants on the quintic 3-fold\label{sec:quintic}}

\subsection{Real Lagrangian submanifolds of quintic 3-folds}

In \cite{Quintic}, the authors investigate open invariants for disks
with boundary on a Lagrangian submanifold $Q_{\mathbb{R}}$ of a quintic
3-fold $Q$. $Q_{\mathbb{R}}$ is obtained as the restriction to $Q$
of the fixed locus of an anti-holomorphic involution on $\mathbb{P}^{4}$;
however, the associated invariants are computed by an integral over
the moduli of stable disk maps to $\mathbb{P}^{4}$, rather than $Q$.
As will be seen in Section \ref{sub:quintic-and-main-result}, the
result of theorem~\ref{thm:main-result} still applies in this setting.
This section reviews the geometric setup of \cite{Quintic}. 

Let $Q\subset\mathbb{P}^{4}$ be a nonsingular quintic hypersurface
with symplectic form $\omega$ obtained from the Fubini-Study metric
on $\mathbb{P}^{4}$. Let $\sigma$ be the anti-holomorphic involution
on $\mathbb{P}^{4}$, described in homogeneous coordinates as 
\[
\sigma\left(\left[z_{0}:z_{1}:z_{2}:z_{3}:z_{4}\right]\right)=\left[\overline{z_{0}}:\overline{z_{2}}:\overline{z_{1}}:\overline{z_{4}}:\overline{z_{3}}\right].
\]
The fixed locus of $\sigma$ is a real Lagrangian submanifold $\mathbb{P}_{\mathbb{R}}^{4}\subset\mathbb{P}^{4}$,
and its restriction to $Q$ is a real Lagrangian submanifold $Q_{\mathbb{R}}\subset Q$. 

For the diagonal $T^{5}$ action on $\mathbb{P}^{4}$, there are five
torus fixed points: 
\begin{align*}
\zeta_{0} & =\left[1:0:0:0:0\right], & \ldots, &  & \zeta_{4} & =\left[0:0:0:0:1\right].
\end{align*}
Only $\zeta_{0}$ is fixed by both the $T^{5}$ action and $\sigma$,
so it is the unique real $T^{5}$-fixed point. Now, consider the rank-2
subtorus $T^{2}\subset T^{5}$ acting by 
\begin{equation}
\left(\theta_{1},\theta_{2}\right)\cdot\left[z_{0}:z_{1}:z_{2}:z_{3}:z_{4}\right]=\left[z_{0}:e^{i\lambda\theta_{1}}z_{1}:e^{-i\lambda\theta_{1}}z_{2}:e^{i\lambda'\theta_{3}}z_{3}:e^{-i\lambda'\theta_{3}}z_{4}\right].\label{eq:PSW-torus-action}
\end{equation}
There are two rational lines invariant with respect to this $T^{2}$
action: $L$, connecting $\zeta_{1}$ to $\zeta_{2}$; and $L'$,
connecting $\zeta_{3}$ to $\zeta_{4}$. 

As in the doubling construction in \cite{KatzLiu}, a holomorphic
degree-$d$ map $\tilde{f}:\mathbb{P}^{1}\rightarrow L$ (or $L$')
can be halved to obtain two winding-$d$ disk maps corresponding to
the two hemispheres of $L$. The boundaries of these disks are the
intersections $L\cap\mathbb{P}_{\mathbb{R}}^{4}$ and $L'\cap\mathbb{P}_{\mathbb{R}}^{4}$.
However, when $d$ is even, not every stable disk map can be obtained
in this way \cite{Quintic}, so only odd-winding disk maps will be
considered in the remainder of this section. 

To enumerate the disk maps $f:\left(\Delta,\partial\Delta\right)\rightarrow\left(Q,Q_{\mathbb{R}}\right)$,
\cite{Quintic} perform an analogous computation to the closed quintic
genus-zero enumeration computation described in \cite{Kontsevich}.
Let $\overline{\mathcal{M}}_{\Delta}\left(\mathbb{P}^{4},\mathbb{P}_{\mathbb{R}}^{4};d\right)$
denote the moduli space of stable disk maps $f:\Delta\rightarrow\mathbb{P}^{4}$
of winding $d$ satisfying $f\left(\partial\Delta\right)\subset\mathbb{P}_{\mathbb{R}}^{4}$,
as in Section~\ref{sub:stable-open-maps}, and let $\hat{F}_{d}$
be the real vector bundle over this moduli space with fiber 
\[
\left.\hat{F}_{d}\right|_{\left[f:\left(\Delta,\partial\Delta\right)\rightarrow\left(\mathbb{P}^{4},\mathbb{P}_{\mathbb{R}}^{4}\right)\right]}=H^{0}\left(C,\tilde{f}^{*}\mathcal{O}_{\mathbb{P}^{4}}\left(5\right)\right)_{\mathbb{R}},
\]
where $\left[\tilde{f}:C\rightarrow\mathbb{P}^{4}\right]$ is the
stable rational map obtained from the stable disk map via reflection,
and the $\mathbb{R}$ subscript denotes real sections. Generically,
the boundary of $\overline{\mathcal{M}}_{\Delta}\left(\mathbb{P}^{4},\mathbb{P}_{\mathbb{R}}^{4};d\right)$
consists of maps with two boundary components. Because $\hat{F}_{d}$
is not trivial near this boundary, the integral $c(\hat{F}_{d})$
over the moduli space is not well-defined. 

To remedy this, consider a generic stable disk map $\left(\Delta,f\right)\in\partial\overline{\mathcal{M}}_{\Delta}\left(\mathbb{P}^{4},\mathbb{P}_{\mathbb{R}}^{4};d\right)$.
$\partial\Delta$ has two components, $B_{1}$ and $B_{2}$. Modifying
$f$ by replacing the image of one of the two components by its image
under the involution $\sigma$ (e.g., defining $f'$ such that $f'\left(B_{2}\right)=\sigma\circ f\left(B_{2}\right)$)
yields another two-component map $\left(f',\Delta\right)$. Let $\widetilde{\mathcal{M}}_{\Delta}\left(\mathbb{P}^{4},\mathbb{P}_{\mathbb{R}}^{4};d\right)$
denote the quotient of $\overline{\mathcal{M}}_{\Delta}\left(\mathbb{P}^{4},\mathbb{P}_{\mathbb{R}}^{4};d\right)$
under the equivalence relation identifying two-component maps of the
form $\left(f,\Delta\right)$ and $\left(f',\Delta\right)$. The vector
bundle $\hat{F}_{d}$ descends to a bundle $F_{d}$ over $\widetilde{\mathcal{M}}_{\Delta}\left(\mathbb{P}^{4},\mathbb{P}_{\mathbb{R}}^{4};d\right)$,
and here \cite{Quintic} prove that the virtual disk count for maps
$f:\left(\Delta,\partial\Delta\right)\rightarrow\left(Q,Q_{\mathbb{R}}\right)$
is given by 
\begin{equation}
N_{d}^{disk}=\int_{\widetilde{M}_{\Delta}\left(\mathbb{P}^{4},\mathbb{P}_{\mathbb{R}}^{4};d\right)}e\left(F_{d}\right)\label{eq:PSW-disk-count}
\end{equation}
\cite[theorem 3]{Quintic}. Localization can be used to compute the
right-hand side of this equation.

\subsection{Disk enumeration and Theorem~\ref{thm:main-result}\label{sub:quintic-and-main-result}}

There are two rational lines $L$, $L'$ fixed by the $T^{2}$ action
on $\mathbb{P}^{4}$ \eqref{eq:PSW-torus-action}. Consequently, there
are four $T^{2}$-fixed disks, identified by their incident torus-fixed
points $\zeta_{i}$. These disk maps are completely characterized
by the winding $p$ and the incident point $\zeta_{i}$. In the notation
of \cite{Quintic}, the disk count is related to the contributions
from each of these disk maps by the formula 
\[
N_{d}^{disk}=\sum_{i=1}^{4}\sum_{p\,\mbox{odd}}\mathrm{Cont}_{\left(\zeta_{i},p\right)}\left(N_{d}^{disk}\right),
\]
where $\mathrm{Cont}_{\left(\zeta_{i},p\right)}\left(N_{d}^{disk}\right)$
denotes the contribution of the disk map $\left(\zeta_{i},p\right)$
to the disk count. The intersection disk term $I\left(\zeta_{i},p\right)$
of $\mathrm{Cont}_{\left(\zeta_{i},p\right)}\left(N_{d}^{disk}\right)$
is the contribution of the unique $T^{2}$-fixed map $f:\left(\Delta,\partial\Delta\right)\rightarrow\left(\mathbb{P}^{4},\mathbb{P}_{\mathbb{R}}^{4}\right)$
with winding $p$ and incident point $\zeta_{i}$. In \cite[lemma 6]{Quintic},
the terms $I\left(\zeta_{i},p\right)$, $i=1,\ldots,4$ are computed
explicitly. 

In fact, this expression can also be obtained using the result of
theorem~\ref{thm:main-result}. For simplicity, restrict attention
to the disk term associated to $I\left(\zeta_{1},p\right)$. Here,
using localization, \cite{Quintic} compute 
\begin{equation}
I\left(\zeta_{1},p\right)=\frac{\left(-1\right)^{\frac{p-1}{2}}}{p}\frac{2\lambda}{p}\frac{\frac{\left(5p\right)!!}{p!p!!}\left(\frac{\lambda}{2p}\right)^{p}}{\prod_{i=0}^{\left(p-1\right)/2}\left(\left(1-\frac{2i}{p}\right)\lambda-\lambda'\right)\left(\left(1-\frac{2i}{p}\right)\lambda+\lambda'\right)},\label{eq:PSW-result}
\end{equation}
where $\lambda$ and $\lambda'$ are the torus weights from \eqref{eq:PSW-torus-action}. 

The vertex of the disk is the point $\zeta_{1}$, so the induced torus
representation $R$ on $T_{0}\Delta$ has weight $\alpha=c_{1}\left(R\right)=2\lambda/p$.
Now, re-write the terms of \eqref{eq:PSW-result} using gamma functions
(recalling the identity $n!!=\pi^{-1/2}2^{\left(n+1\right)/2}\Gamma\left(\frac{1}{2}n+1\right)$
for $n$ odd): 
\begin{eqnarray*}
\frac{\left(5p\right)!!}{p!p!!}\left(\frac{\lambda}{2p}\right)^{p} & = & \left(\frac{2\lambda}{p}\right)^{p}\frac{\Gamma\left(\frac{5p}{2}+1\right)}{\Gamma\left(p+1\right)\Gamma\left(\frac{p}{2}+1\right)},\\
\prod_{i=0}^{\left(p-1\right)/2}\left(1-\frac{2i}{p}\right)\lambda-\lambda' & = & \left(\frac{2\lambda}{p}\right)^{\left(p+1\right)/2}\frac{\Gamma\left(1+\frac{p}{2\lambda}\left(\lambda-\lambda'\right)\right)}{\Gamma\left(\frac{1}{2}-\frac{\lambda'p}{2\lambda}\right)},\\
\prod_{i=0}^{\left(p-1\right)/2}\left(1-\frac{2i}{p}\right)\lambda+\lambda' & = & \left(\frac{2\lambda}{p}\right)^{\left(p+1\right)/2}\frac{\Gamma\left(1+\frac{p}{2\lambda}\left(\lambda+\lambda'\right)\right)}{\Gamma\left(\frac{1}{2}+\frac{\lambda'p}{2\lambda}\right)}.
\end{eqnarray*}
Substituting these expressions into the third fraction of \eqref{eq:PSW-result}
yields 
\[
\frac{\Gamma\left(\frac{5p}{2}+1\right)\Gamma\left(\frac{1}{2}-\frac{\lambda'p}{2\lambda}\right)\Gamma\left(\frac{1}{2}+\frac{\lambda'p}{2\lambda}\right)}{\left(\frac{2\lambda}{p}\right)\Gamma\left(p+1\right)\Gamma\left(\frac{p}{2}+1\right)\Gamma\left(1+\frac{p}{2\lambda}\left(\lambda-\lambda'\right)\right)\Gamma\left(1+\frac{p}{2\lambda}\left(\lambda+\lambda'\right)\right)}
\]
\[
=\left(\frac{\pi}{z\sin\left(\pi\left(\frac{1}{2}+\frac{\lambda'}{z}\right)\right)}\right)\left(\frac{\Gamma\left(\frac{5\lambda}{z}+1\right)}{\Gamma\left(\frac{2\lambda}{z}+1\right)\Gamma\left(\frac{\lambda}{z}+1\right)\Gamma\left(\frac{\lambda-\lambda'}{z}+1\right)\Gamma\left(\frac{\lambda+\lambda'}{z}+1\right)}\right)
\]
after the additional substitution $z=\alpha=2\lambda/p$. 

This formula now closely resembles \eqref{eq:disk-term-Delta}. By
definition, the gamma class is multiplicative. Hence, $\hat{\Gamma}_{Q}$
can be obtained from the normal bundle exact sequence of sheaves 
\[
\xymatrix{0\ar[r] & T_{Q}\ar[r] & T_{\mathbb{P}^{4}}\ar[r] & \mathcal{O}_{\mathbb{P}^{4}}\left(5\right)\ar[r] & 0}
,
\]
yielding 
\[
\hat{\Gamma}_{Q}=\frac{\hat{\Gamma}_{\mathbb{P}^{4}}}{\hat{\Gamma}_{\mathcal{O}\left(5\right)}}.
\]
An additional subtlety is that the above exact sequence is not torus-equivariant
for the real quintic. However, from the hard Lefshetz theorem, it
suffices to write the (non-equivariant) gamma class as a ratio, and
then choose an equivariant lift for each of the factors. 

The weights of the tangent space $T_{\mathbb{P}^{4}}$ at the fixed
point $\zeta_{1}$ are $2\lambda$, $\lambda$, $\lambda+\lambda'$,
and $\lambda-\lambda'$, and the bundle $\mathcal{O}_{\mathbb{P}^{4}}\left(5\right)$
has weight $5\lambda$ at this fixed point. Therefore, the second
term in the product above is the localized form of the inverse of
$\hat{\Gamma}_{Q}$. Finally, because $p=2\lambda/z$, $\left(-1\right)^{\left(p-1\right)/2}=\sin\left(\pi\frac{\lambda}{z}\right)$.
Combining these observations, \eqref{eq:PSW-result} can be expressed
as 
\begin{equation}
I\left(\zeta_{1},p\right)=\frac{\pi}{p\hat{\Gamma}_{Q}\sin\left(\pi\frac{\lambda+\lambda'}{z}\right)},\label{eq:gamma-PSW}
\end{equation}
which has a similar form to \eqref{eq:disk-term-Delta}. The disk
is contained in the rational line $L$. In local coordinates 
\[
\left(x_{1},x_{2},x_{3},x_{4}\right)=\left(\frac{z_{0}}{z_{1}},\frac{z_{2}}{z_{1}},\frac{z_{3}}{z_{1}},\frac{z_{4}}{z_{1}}\right)
\]
near the vertex $\zeta_{1}$, the disk map is $t\mapsto\left(0,t^{p},0,0\right)$,
and $\lambda+\lambda'$ is the weight associated to a normal direction
to the disk (namely, $\partial_{x_{4}}$). The only discrepancy between
\eqref{eq:gamma-PSW} and \eqref{eq:disk-term-Delta} is an additional
factor of $z$ in the latter equation. As these are invariants appearing
from the localization of an integral \eqref{eq:PSW-disk-count} over
$\widetilde{\mathcal{M}}_{\Delta}\left(\mathbb{P}^{4},\mathbb{P}_{\mathbb{R}}^{4};d\right)$,
rather than $\overline{\mathcal{M}}_{\Delta}\left(Q,Q_{\mathbb{R}};d\right)$,
there is no a priori reason to expect a form of theorem~\ref{thm:main-result}
to apply.

\section{Lagrangian Cycles in Large $N$ Duality\label{sec:torus-knot-lagrangians}}

\subsection{Lagrangian cycles and the conifold transition}

In addition to Lagrangians appearing as the fixed loci of anti-holomorphic
involutions, there is another family of Lagrangians on $X=\mathcal{O}_{\mathbb{P}^{1}}\left(-1,-1\right)$
motivated by large $N$ duality and knot theory. Recent work in this
area has yielded many connections between knot theory and Gromov-Witten
theory (\cite{BriniEynardMarino,DSV,GuJockersKlemmSoroush}); this
section reviews the geometric relationship between knots on $S^{3}$
and open Gromov-Witten theory on $X$.

Recall that $X$ can be identified with the resolved conifold---$X$
is the small resolution of the conifold singularity 
\[
xy-zw=0
\]
in $\mathbb{C}^{4}$. In particular, by blowing up the subspace $y=z=0$,
$X$ can be described by the equations 
\begin{align*}
xy-zw & =0, & x\lambda & =w\rho, & y\lambda & =z\rho,
\end{align*}
where $\left(x,y,z,w\right)\in\mathbb{C}^{4}$ and $\left[\lambda:\rho\right]\in\mathbb{P}^{1}$.
The conifold singularity is also the singular limit of the smooth
hypersurface threefold $Y_{\mu}\subset\mathbb{C}^{4}$ defined by
\[
xy-zw=\mu,
\]
where $\mu\in\mathbb{R}_{>0}$. As described in \cite{DSV}, $Y_{\mu}$
is symplectomorphic to the cotangent bundle $T_{S^{3}}^{*}$. The
base $S_{\mu}\cong S^{3}$ is the fixed locus of the anti-holomorphic
involution $\sigma\left(x,y,z,w\right)=\left(\overline{z},-\overline{w},\overline{x},-\overline{y}\right)$,
expressed by the equations $\left|x\right|^{2}+\left|y\right|^{2}=\mu$. 

The large $N$ duality conjecture states that the large $N$ limit
of the topological A-model on $Y_{\mu}$ with $N$ Lagrangian branes
wrapping $S_{\mu}$ is equivalent the topological A-model on $X$
\cite{GopakumarVafa}. This has been checked in several ways. First,
according to \cite{WittenCSThy}, the topological A-model on $Y_{\mu}$
with $N$ Lagrangian branes wrapping $S_{\mu}$ is equivalent to the
$U\left(N\right)$ Chern-Simons theory on $S_{\mu}$. Then, in the
large $N$ expansion, the partition function $Z_{CS}\left(k,N\right)$
is equivalent to the topological A-model partition function $Z_{X}\left(g_{s},t\right)$
\cite{GopakumarVafa}. The parameters determining the A-model theory
on $X$ are the string coupling constant $g_{s}$ and the symplectic
area $t$ of the zero section $\mathbb{P}^{1}\subset X$, which are
related to the Chern-Simons parameters $k$ and $N$ by 
\begin{align*}
g_{s} & =\frac{2\pi}{k+N}, & t & =-\frac{2\pi iN}{k+N}.
\end{align*}

\begin{figure}[th]
\begin{center}
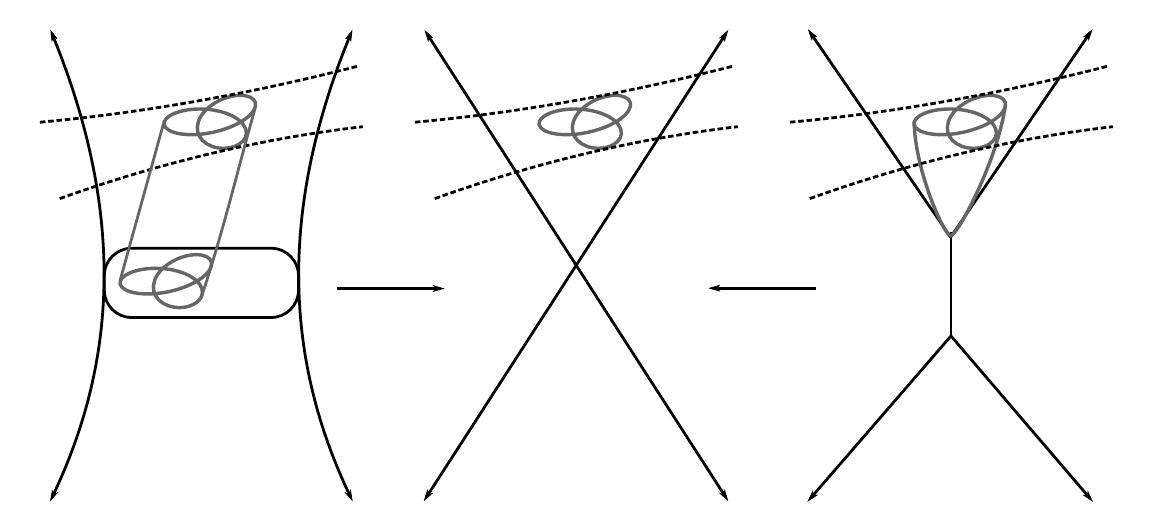
\end{center}

\caption{\label{fig:conifold-transition}The conifold transition for lifted
Lagrangian cycles.}

\begin{minipage}[t]{1\columnwidth}%
This figure depicts the conifold transition. The Lagrangian $\tilde{L}\subset Y_{\mu}\cong T_{S^{3}}^{*}$
is constructed by shifting the conormal bundle of a knot $K\subset S^{3}$
off of the zero section. This lift introduces a holomorphic cylinder
$C$ connecting the knot on $S^{3}$ to its image in $\tilde{L}$.
$Y_{0}$ is the conifold singularity $xz-yw=0$ in $\mathbb{C}^{4}$.
The map $\phi_{\mu}:Y_{\mu}\rightarrow Y_{0}$ is a symplectomorphism
away from the zero section, so $\phi_{\mu}\left(\tilde{L}\right)$
is a Lagrangian submanifold of $Y_{0}$. $X\cong\mathcal{O}_{\mathbb{P}^{1}}\left(-1,-1\right)$
is the small resolution of the conifold singularity, and $\sigma_{\epsilon}:X\rightarrow Y_{0}$
is the corresponding natural map. In fact, there are a family of such
symplectomorphisms, where $\epsilon$ parametrizes the symplectic
form on the zero section $\mathbb{P}^{1}\subset X$. Hence, $L:=\sigma_{\epsilon}^{-1}\circ\phi_{\mu}\left(\tilde{L}\right)$
is a Lagrangian submanifold of $X$. The holomorphic disk $D$ is
the image of $C$ under the conifold transition. %
\end{minipage}
\end{figure}

Large $N$ duality is extended to incorporate Wilson loops in \cite{OoguriVafa}.
Following \cite{WittenCSandHOMFLY}, Wilson loop observables in the
Chern-Simons theory on $S^{3}$ correspond to colored HOMFLY polynomials
of knots $K\subset S^{3}$. The conormal bundle $N_{K}^{*}$ to a
knot $K\subset S^{3}$ is Lagrangian submanifold of $T_{S^{3}}^{*}$.
The main difficulty in extending large $N$ duality in this manner
is determining the corresponding A-model on $X$: $N_{K}^{*}$ intersects
the zero section $S^{3}$ in the knot $K$, which becomes contracted
after the conifold transition. To remedy this difficulty, the Lagrangian
cycle $N_{K}^{*}$ must be lifted to a new Lagrangian $\tilde{L}$
disjoint from the zero section before performing the conifold transition
\cite{AMV,MarinoVafa}, as depicted in Figure~\ref{fig:conifold-transition}.

Such a lift is easy to construct: define coordinates $\left(\overrightarrow{u},\overrightarrow{v}\right)$
for $T_{S^{3}}^{*}$ by 
\[
T_{S^{3}}^{*}=\left\{ \left(\overrightarrow{u},\overrightarrow{v}\right)\in\mathbb{R}^{4}\times\mathbb{R}^{4}:\left|\overrightarrow{u}\right|=1,\overrightarrow{u}\cdot\overrightarrow{v}=0\right\} .
\]
 Any knot $K\subset S^{3}$ is given by a parametrization $\overrightarrow{u}=f\left(\theta\right)$.
Then, the conormal bundle $N_{K}^{*}$ can be expressed as 
\[
N_{K}^{*}=\left\{ \left(\overrightarrow{u},\overrightarrow{v}\right)\in T^{*}S^{3}:\overrightarrow{u}=f\left(\theta\right),\frac{df}{d\theta}\cdot\overrightarrow{v}=0\right\} .
\]
Lifts of $N_{K}^{*}$ can be specified by maps $g:S^{1}\rightarrow T_{f\left(\theta\right)}^{*}S^{3}$
such that $\frac{df}{d\theta}\cdot g\left(\theta\right)\neq0$: for
such a $g$, define the lifted conormal bundle $\tilde{L}$ to be
\[
\tilde{L}:=\left\{ \left(\overrightarrow{u},\overrightarrow{v}\right)\in T^{*}S^{3}:\overrightarrow{u}=f\left(\theta\right),\frac{df}{d\theta}\cdot\left(\overrightarrow{v}-g\left(\theta\right)\right)=0\right\} .
\]
The image of $\tilde{L}$ under the conifold transition will be a
Lagrangian $L\subset X$, and the open A-model on $X$ with this Lagrangian
boundary can be computed. Shifting $N_{K}^{*}$ off of the zero section
modifies large $N$ duality in the following ways: The lift of $N_{K}^{*}$
to $\tilde{L}$ introduces corrections to the Wilson loop observables
in the Chern-Simons theory proportional to the area of the holomorphic
cylinder $C$ connecting the lift of the knot to its image in the
zero section \cite{DSV}. Instead of the closed A-model on $X$, the
corresponding theory should be an open A-model with Lagrangian boundary
$L$. This statement of large $N$ duality is found to be true for
torus knots in \cite{DSV}, and their construction provides a novel
source of Lagrangians.

\subsection{Toric Lagrangian cycles and Theorem~\ref{thm:main-result}\label{sub:DSV-comparison}}

\begin{figure}[th]
\begin{center} 
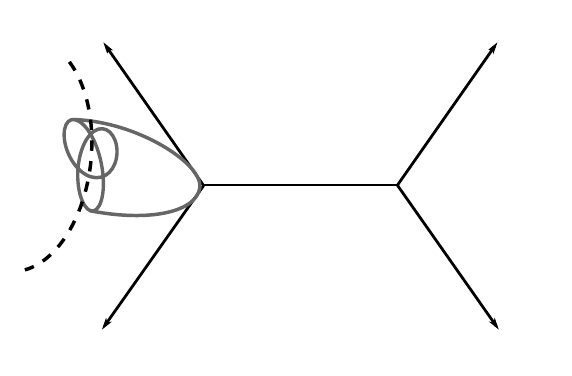
\end{center}

\caption{\label{fig:DSV-Lagrangian}A moment polytope picture of torus knot
Lagrangians.}

\begin{minipage}[t]{1\columnwidth}%
This figure depicts the geometry described in Section~\ref{sub:DSV-comparison}.
The Lagrangian $L$ is the image of a shifted conormal bundle to a
knot in $S^{3}$ under the conifold transition. Local coordinates
$\left(x,y\right)$ and $\left(z,w\right)$ parametrize the fibers
of $X$ in a neighborhood of the two vertices of the moment polytope.
The boundary of the disk $D$ is symplectomorphic to the torus knot
(in the depiction above, the trefoil). $D$ is contained entirely
in the $x$-$y$ face of the polytope, and the disk map can be written
in local coordinates as $t\mapsto\left(\xi=0,x=b_{1}^{s}t^{ws},y=b_{1}^{r}t^{wr}\right)$. %
\end{minipage}
\end{figure}

It is important to note that the Lagrangians considered in \cite{DSV}
are not obtained as the fixed loci of anti-holomorphic involutions,
so there is no a priori reason to expect that the formula proposed
in theorem~\ref{thm:main-result} should apply in this situation.
For the $\left(r,s\right)$ torus knot, the corresponding Lagrangian
$L$ is found to be fixed under the torus action 
\[
\rho_{\theta}\left(\left(x,y,z,w\right),\left[\lambda:\rho\right]\right)=\left(e^{is\theta}x,e^{ir\theta}y,e^{-is\theta}z,e^{-ir\theta}w\right),\left[e^{-i\left(r+s\right)\theta}\lambda:\rho\right].
\]
There is only one holomorphic disk in $X$ fixed by this $S^{1}$
action, and it lies entirely in the $x$-$y$ face of the moment polytope,
as depicted in Fig.~\ref{fig:DSV-Lagrangian}. 
 A neighborhood of the disk can be described by local coordinates
$x$, $y$, $\xi=\lambda/\rho$. In these coordinates, the disk map
is 
\[
t\mapsto\left(\xi=0,x=b_{1}^{s}t^{ws},y=b_{1}^{r}t^{wr}\right),
\]
where $\left|t\right|\leq1$ and $b_{1}\in\mathbb{R}_{>0}$ is a constant
obtained from the geometric construction in \cite{DSV}. After a lengthy
localization calculation, \cite{DSV} compute the winding-1 open Gromov-Witten
invariants with Lagrangian boundary $L$. This computation readily
generalizes to higher winding \cite{GuJockersKlemmSoroush}, and gives
the following expression for $D_{X,L}$: 
\begin{equation}
D_{X,L}=\left(-1\right)^{ws}\frac{\prod_{k=1}^{ws-1}\left(r+s-\frac{k}{w}\right)}{w\prod_{k=0}^{ws-1}\left(s-\frac{k}{w}\right)}.\label{eq:DSV-higher-winding}
\end{equation}
This can be re-written in terms of gamma functions in the following
way: 
\[
\frac{\prod_{k=1}^{ws-1}\left(r+s-\frac{k}{w}\right)}{\prod_{k=0}^{ws-1}\left(s-\frac{k}{w}\right)}=\frac{\Gamma\left(wr+ws\right)}{\left(\frac{1}{w}\right)\Gamma\left(ws+1\right)\Gamma\left(wr+1\right)}.
\]
Locally (Figure~\ref{fig:DSV-Lagrangian}), the weights of the torus
action are $\lambda_{\xi}=-r-s$, $\lambda_{x}=s$, $\lambda_{y}=r$.
The induced torus action on $\Delta$ is $t\mapsto e^{i\theta/w}t$,
so $\alpha=\frac{1}{w}$. Replacing $\alpha=z$ and substituting these
weights into the above formula yields 
\[
=\frac{\Gamma\left(-\frac{\lambda_{\xi}}{z}\right)}{wz\Gamma\left(\frac{\lambda_{x}}{z}+1\right)\Gamma\left(\frac{\lambda_{y}}{z}+1\right)}=\frac{\pi}{wz\widehat{\Gamma}_{X}\sin\left(\pi\frac{\lambda_{\xi}}{z}\right)}
\]
after Euler's reflection identity. As remarked above, the Lagrangian
$L$ is not the fixed locus of an anti-holomorphic involution. However,
the result still applies. Let $C$ be the $S^{1}$-invariant curve
given locally by $\left(0,t^{s},t^{r}\right)$ for $t\in\mathbb{C}$.
Then, $L\cap C=\partial D$ and $\lambda_{\xi}$ is the weight of
an $S^{1}$-invariant normal direction to $C$ as $C$ is entirely
contained in the $x$-$y$ plane of the moment polytope. The author
finds it curious that the result of theorem~\ref{thm:main-result}
appears to apply in this situation, and hopes that this is evidence
that, properly formulated, a more general version of this theorem
exists.

\end{document}